\documentclass[manuscript]{aastex}

\shorttitle{Hectospec}
\shortauthors{Fabricant et al.}

\begin{document}

\title{Hectospec, the MMT's 300 Optical Fiber-Fed Spectrograph}

\author{Daniel Fabricant, Robert Fata, John Roll, Edward Hertz, Nelson
Caldwell, Thomas Gauron, John Geary, Brian McLeod, Andrew Szentgyorgyi, Joseph
Zajac, Michael Kurtz, Jack Barberis, Henry Bergner, Warren Brown, Maureen Conroy, Roger Eng,
Margaret Geller, Richard Goddard, Mike Honsa, Mark Mueller,
Douglas Mink, Mark Ordway, Susan Tokarz, Deborah Woods, William Wyatt}

\affil{Smithsonian Astrophysical Observatory, a member of the Harvard-Smithsonian Center for Astrophysics, Cambridge, MA 02138}

\email{dfabricant@cfa.harvard.edu}

\author{Harland Epps}
\affil{Lick Observatory, UC Santa Cruz, Santa Cruz, CA 95064}

\and

\author{Ian Dell'Antonio}
\affil{Brown University, Box 1843, Providence, RI 02912}

\begin{abstract}

The Hectospec is a 300 optical fiber fed spectrograph commissioned at the MMT
in the spring of 2004. In the configuration pioneered by the {\it Autofib}
instrument at the AAT, Hectospec's fiber probes are arranged in a radial, 
``fisherman on the pond" geometry and held in position with small magnets. A
pair of high-speed six-axis robots move the 300 fiber buttons between
observing configurations within $\sim$300 s and to an accuracy of $\sim$25
$\mu$m. The optical fibers run for 26 m between the MMT's focal surface and
the bench spectrograph operating at R$\sim$1000-2000. Another high dispersion
bench spectrograph offering R$\sim$35,000, Hectochelle, is also available. The
system throughput, including all losses in the telescope optics, fibers, and
spectrograph peaks at $\sim$10\% at the grating blaze in 1$^{\prime\prime}$
FWHM seeing. Correcting for aperture losses at the 1.5$^{\prime\prime}$
diameter fiber entrance aperture, the system throughput peaks at $\sim$17\%,
close to our prediction of 20\%. Hectospec has proven to be a workhorse
instrument at the MMT. Hectospec and Hectochelle together were scheduled for
1/3 of the available nights since its commissioning. Hectospec has returned
$\sim$60,000 reduced spectra for 16 scientific programs during its first year
of operation. 

\end{abstract}


\section{Introduction}

\subsection{Hectospec Overview}

The Hectospec is a powerful fiber-fed spectrograph at the MMT Observatory in
routine operation since April 2004. During its first year of operation,
Hectospec obtained 60,000 spectra during 79 scheduled nights, with 48 nights
of clear weather.  Hectochelle, Hectospec's high dispersion partner
spectrograph, was scheduled for an additional 31 nights. In total, Hectospec
and Hectochelle were scheduled for 1/3 of the available MMT nights. During
Hectospec's first year, it served 16 scientific programs.

\citet{fab94} and \citet{fab98} outline the basic design of the Hectospec
fiber positioner, its bench spectrograph, and optical fiber probes. Here, we
emphasize: (1) the design refinements that occurred between 1998 and
Hectospec's delivery to the telescope in August 2003 and (2) Hectospec's
performance at the telescope.

\subsection{The Converted MMT and Its Optics}

The MMT's f/5 optical system is comprised of a 6.5 m primary mirror, a 1.7 m
secondary mirror, a series of large telescope baffles, a large refractive
corrector, and a wave front sensor used periodically to correct the figure of
the primary mirror and the telescope collimation. \citet{fab04} contains an
overview of these components. \citet{fata94} and \citet{cal04} describe the
f/5 secondary and its support. \citet{fata93} and \citet{fata04} discuss the
f/5 wide field corrector in more detail. \citet{pick04} describe the
Shack-Hartmann f/5 wave front sensor.

The wide field corrector has two modes of operation. For wide field
spectroscopy with optical fibers, the corrector provides a 1$^{\circ}$
diameter field of view on a curved but telecentric focal surface 0.6 meters in
diameter. The sag of the spectroscopic focal surface is 8 mm. In the imaging
mode, the corrector provides a field of view 0.58$^{\circ}$ in diameter on a
flat focal surface. The spectroscopic mode of the corrector includes
counterrotating atmospheric dispersion compensation prisms that are removed
for the imaging mode. 

The commissioning of Hectospec required the prior installation of the 1.7 meter
f/5 secondary mirror at the MMT, which took place in April 2003. The wide
field corrector and the f/5 wave front sensor were commissioned during May and
June 2003. The first commissioning run with Hectospec took place in October
2003, and during this run we obtained observations of nearby
galaxy clusters. The second Hectospec commissioning run was scheduled for
April 2004. In mid April 2004 Hectospec was placed into routine operation. 

\subsection{Hectochelle}

Hectochelle uses Hectospec's robotic fiber positioner and optical fibers to
feed a high dispersion bench spectrograph in place of Hectospec's moderate
dispersion spectrograph. Single orders, or several overlapping orders,
dispersed by Hectochelle's echelle grating are selected with bandpass filters.
\citet{saint98} describe Hectochelle's design.

\section{Hectospec Fiber Positioner}

\subsection{Introduction}

Hectospec's robotic fiber positioner is unique for its high fiber placement speed
while retaining high positioning accuracy. The high speed robots meet their
design goals reliably: 150 paired moves of the tandem robots in under 300
seconds, and with 25 $\mu$m fiber positioning accuracy. The fiber robots attain
this combination of high speed and accuracy due to a very stiff mechanical
system that settles rapidly, powerful servo motors that rapidly accelerate the
robots, and a sophisticated control system that incorporates many safety
features while minimizing communication overhead between system components.

Hectospec's fiber geometry was first used by {\it Autofib} \citep{parry86}, an
early robotic fiber-fed spectrograph for the Anglo-Australian Telescope with a
single Cartesian robot positioning 64 optical fibers. Hectospec's 300 fiber
buttons are held onto the (400 series stainless) steel focal surface with
NdFeB magnets mounted in the bottom of the fiber button. The magnets provide
~200 g of holding force normal to the focal surface and $>$50 g of holding
force in the plane of the focal surface.

\subsection{Fiber Buttons and Focal Surface Assembly}

\begin{figure}
\epsscale{.5}
\plotone{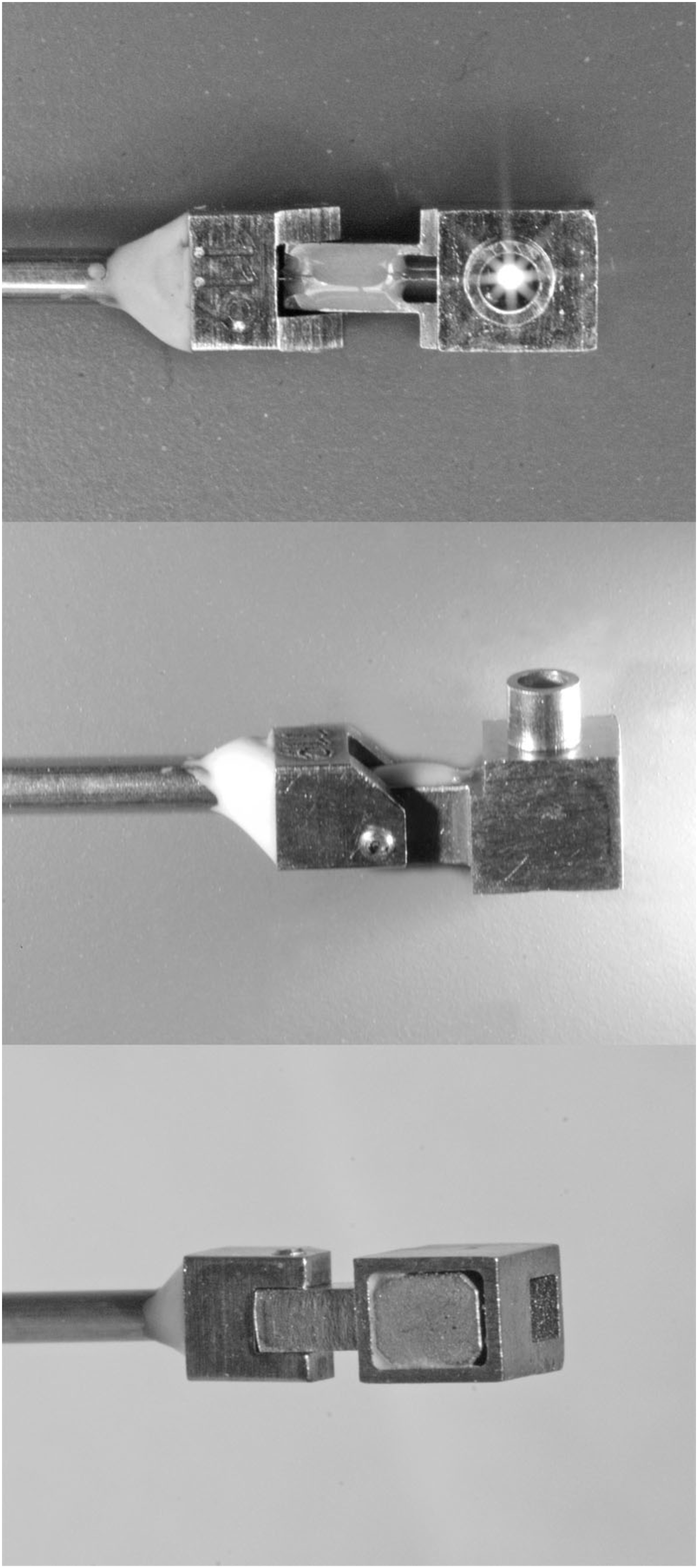}
\caption{Three views of the fiber button assembly. Light from the backlit fiber is visible in the upper panel.  The pivot point that allows the button assembly to accommodate the curved focal surface is visible in the middle panel.  The rectangular NeFeB magnet bonded into the button assembly is visible in the lower panel.\label{fig1}}
\end{figure}

\begin{figure}
\epsscale{1}
\plotone{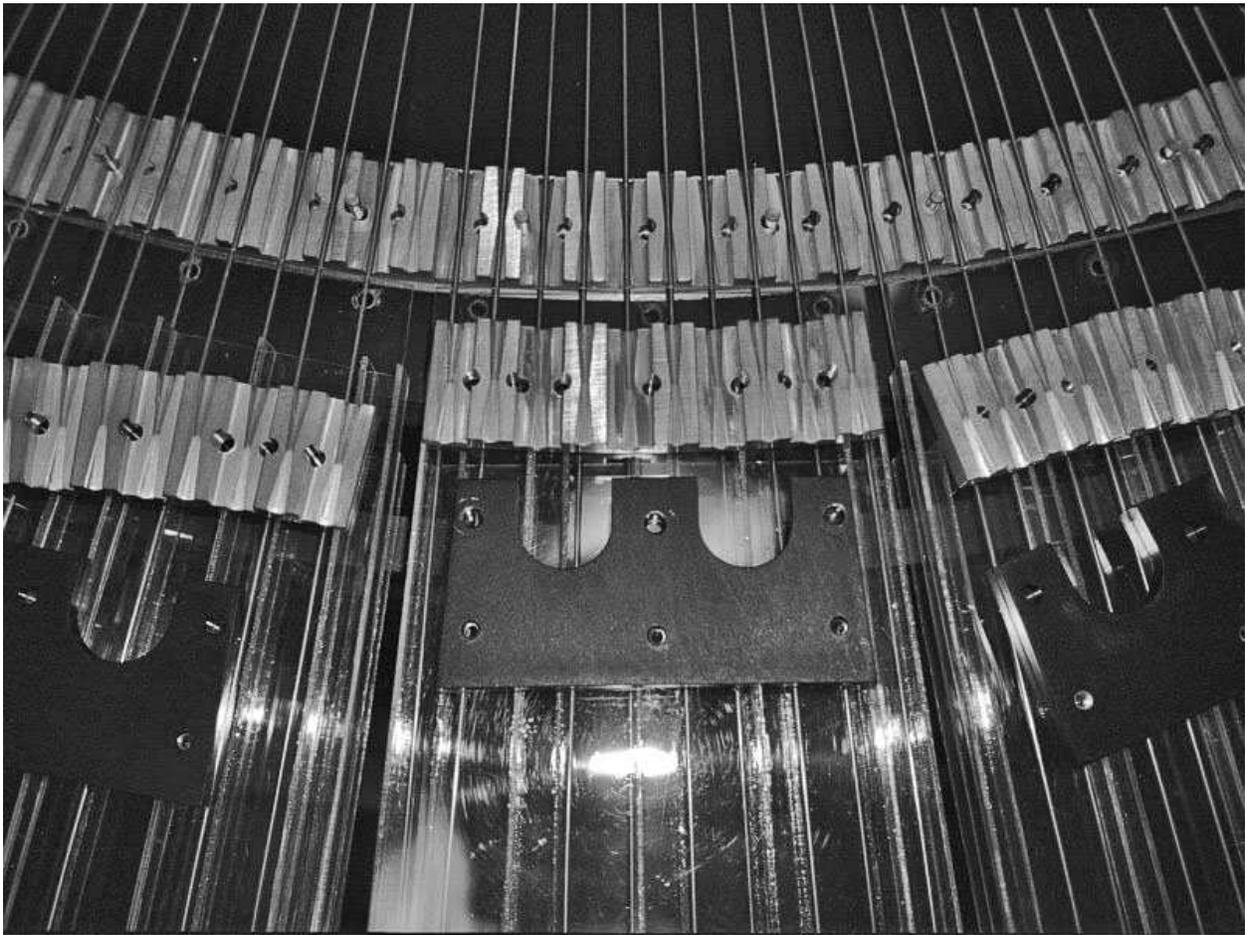}
\caption{Disassembled fiber pivots. Pins register and screws hold the upper and lower halves of the fiber pivots together when the pivots are assembled.  The pivot points are defined by the intersection of two cones. The pivots are
arranged in two vertical levels so that the fiber syringes are fed into two
vertical levels of separator trays.\label{fig2}}
\end{figure}

Figure \ref{fig1} shows a fiber button assembly from three angles. The 1.27 mm
diameter stainless steel tube that protects the fiber after its exit from the
fiber button is supported on its far end in brass pivot blocks. The fiber
pivots are located 430 mm from the focal plane's center. The pivots are
arranged in two vertical levels so that the fiber syringes are fed into two
vertical levels of separator trays. Using two levels provides more horizontal
space in the separator trays and protects the syringes from excessive bending
as the fiber button is moved tangentially on the focal surface. Figure
\ref{fig2} shows the (disassembled) fiber pivots outside the edge of the
focal surface. 

\subsection{Mechanical Design}

\begin{figure}
\epsscale{1}
\plotone{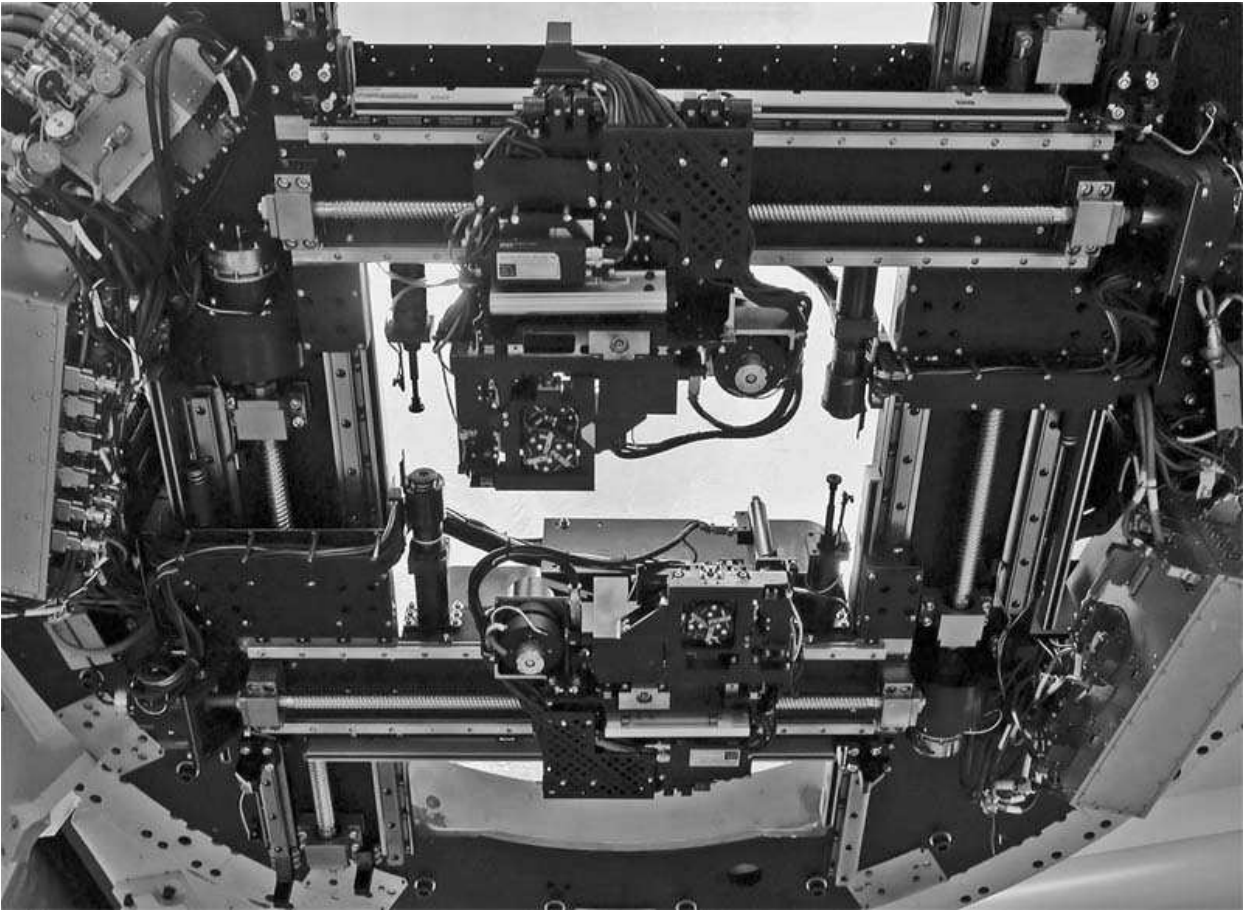}
\caption{Upper unit of the fiber positioner.  The X-axis stages and ball screws run vertically and the Y-axis stages and ballscrews run horizontally in this picture.  The Z-axes run into the paper.  The $\Theta$ gimbal axis tilts along the X direction, while the $\Phi$ axis tilts along the Y direction.  The two grippers are the two circular assemblies near the center of the picture.  The pair of X-axis collision bumpers are visible above and to the side of the two Z-axis assemblies.\label{fig3}}
\end{figure}

\begin{figure}
\epsscale{1}
\plotone{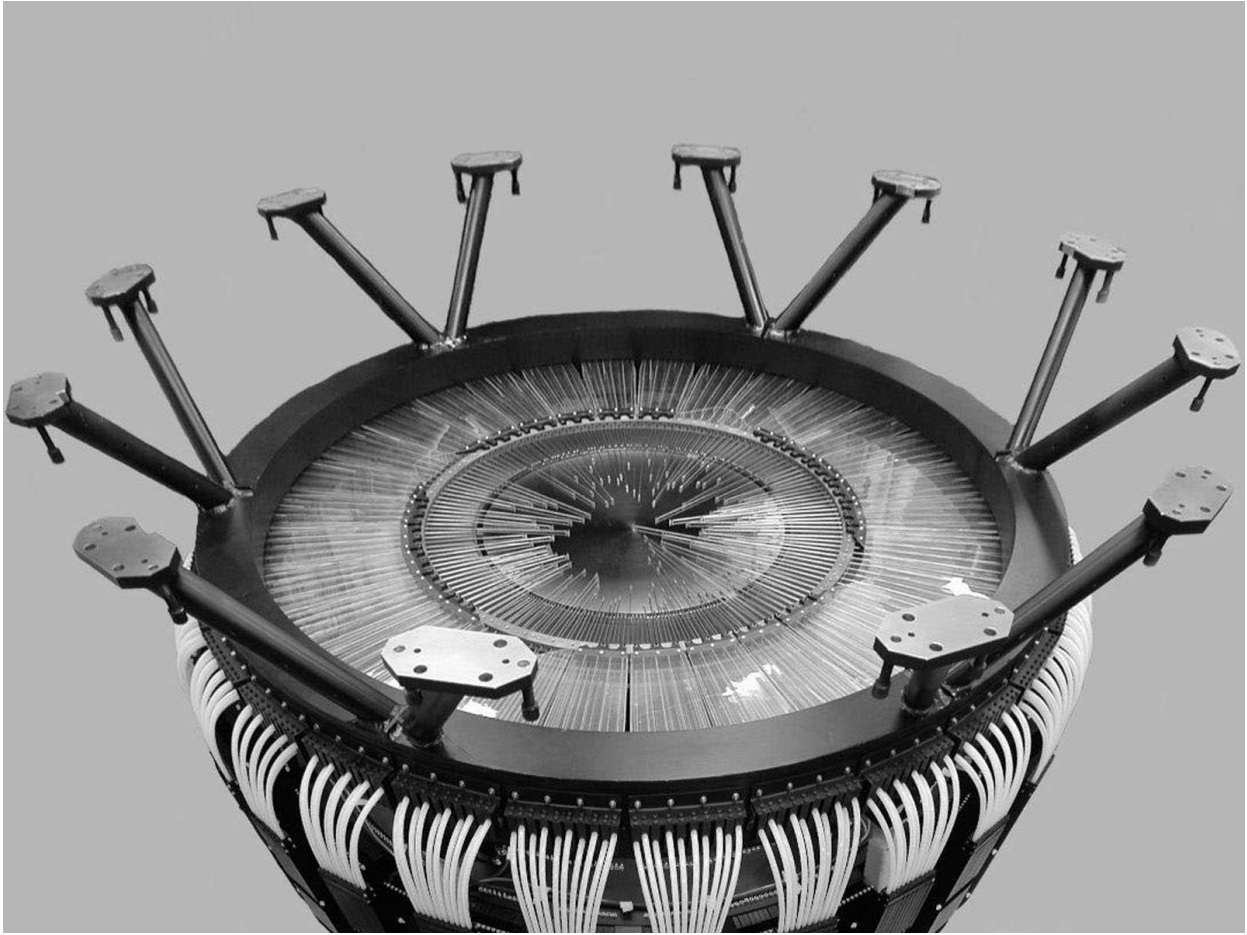}
\caption{Lower unit of the fiber positioner.  This section of the fiber positioner can be unbolted from the upper unit that contains the robots (as shown).  The lower unit contains the 300 fiber optic probes, the curved focal plate that holds the fiber optic probes, guide trays for the fiber probes, as well as the three guider probes (not visible).\label{fig4}}
\end{figure}

\begin{figure}
\epsscale{1}
\plotone{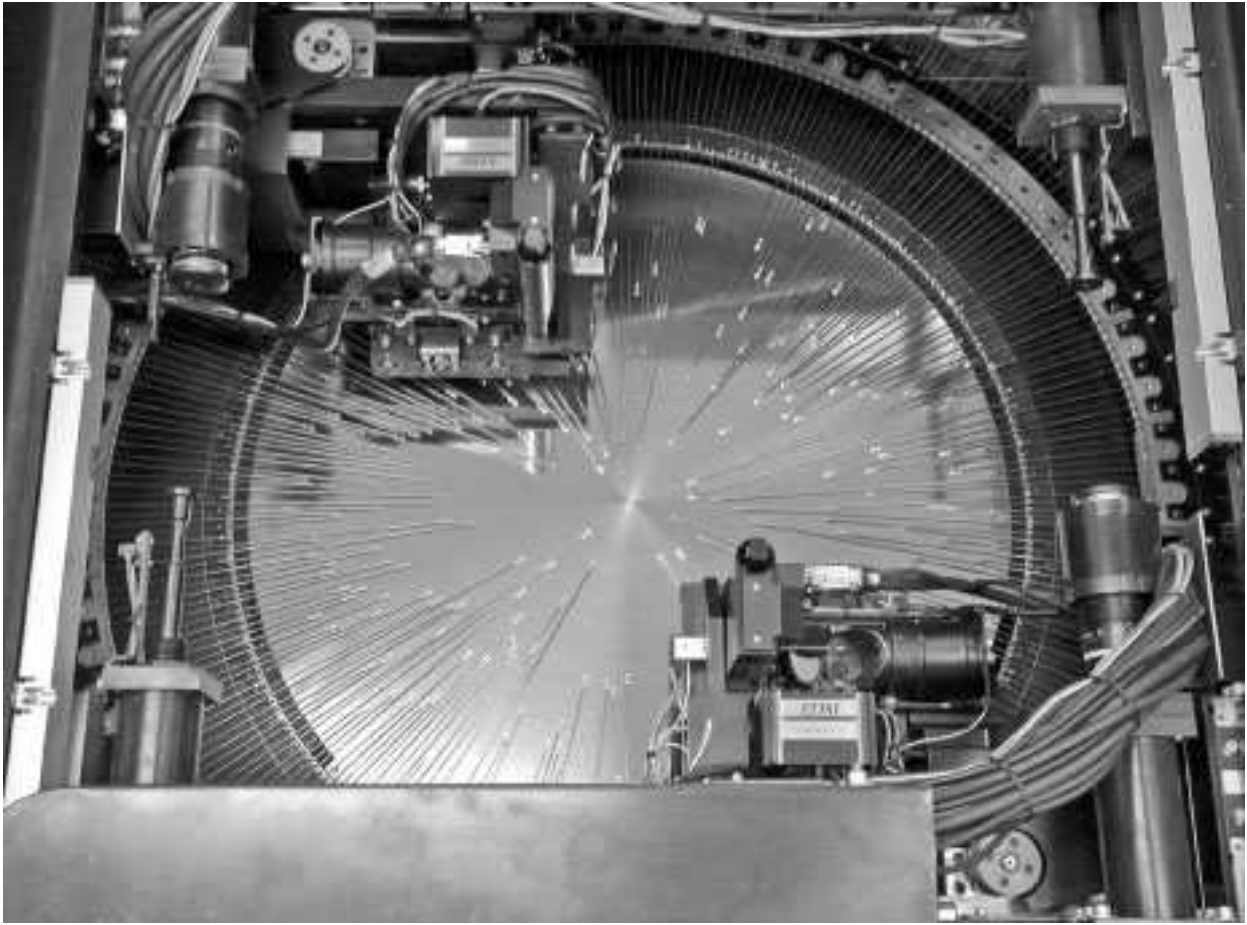}
\caption{Top view of the assembled fiber positioner.  This photograph was taken with upper cover assembly removed to allow a clearer view of the two six-axis robots.  The X-axis collision bumpers are visible at the left and right.  The two rectangular assemblies marked ``EOSI" are the intensified TV cameras that travel with the robots.  These TV cameras are used for calibration, alignment tests, and the initial acquistion of guide stars.\label{fig5}}
\end{figure}

The fiber positioner can be separated into two parts to allow servicing of the
robots and optical fibers. The upper unit (Figure \ref{fig3}) contains the two
six axis robots and most of the electronics. The lower unit (Figure
\ref{fig4}) contains the fiber probes, the fiber shelves (to prevent tangling
of the optical fibers), the three guider probes and their track, the
intensified camera for the guider probes, and the fiber derotator assembly
that allows the fibers to follow instrument rotation. The two units are
separated by removing protective covers and then unbolting the upper end of
the struts that connect the two assemblies. Figure \ref{fig5} is a top view of
the assembled upper and lower units.

\begin{figure}
\epsscale{1}
\plotone{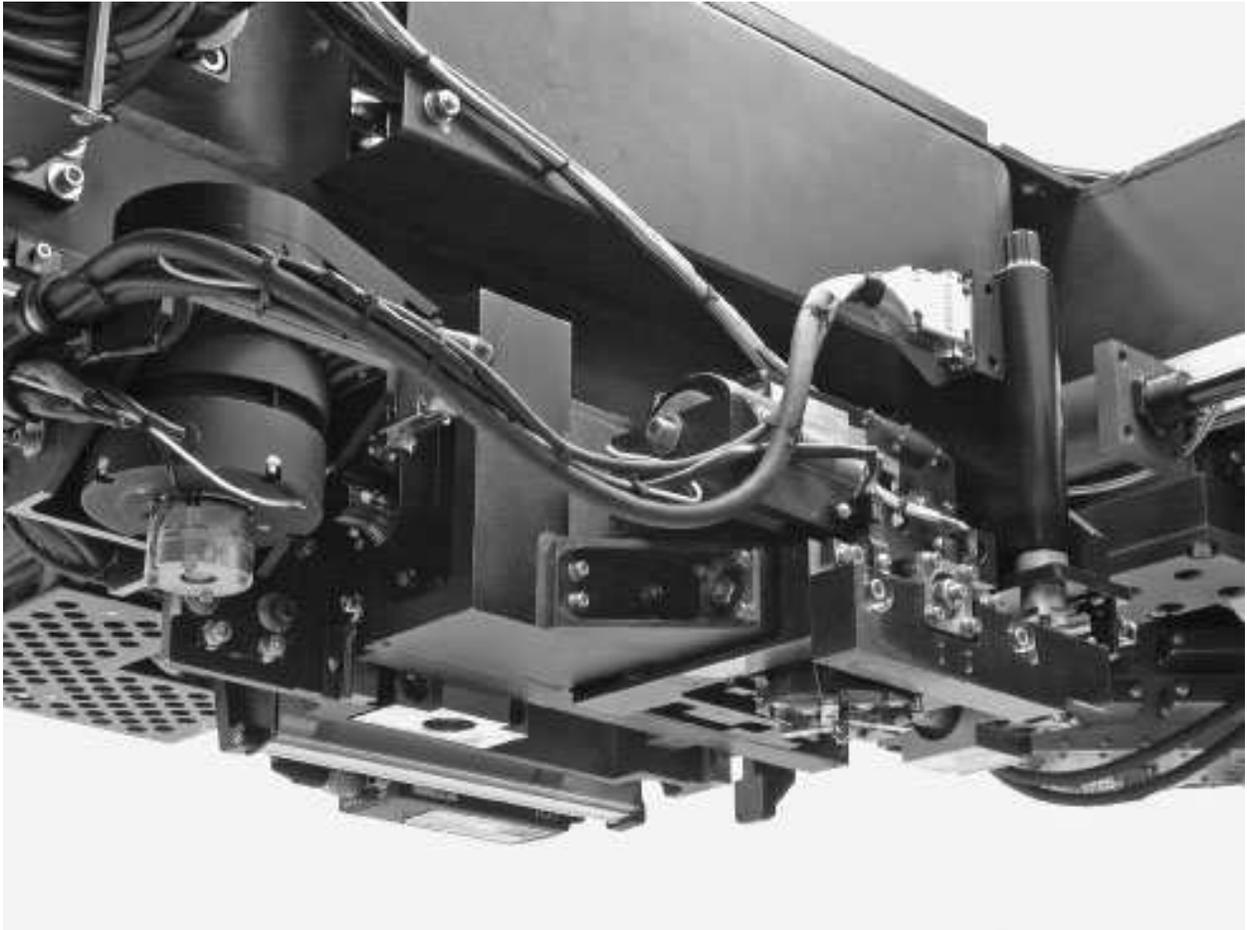}
\caption{Z axis assembly of the fiber positioner.  The Z axis motor, encoder and brake are visible at the center left.  The cover for the metal drive band is mounted to the top of the Z axis motor assembly.  The gripper jaws are visible at the lower right.  The $\Theta$ axis actuator is located at the center right.  The bearing assemblies for the $\Theta$ and $\Phi$ axes are located just above the gripper jaws at the lower right.  The corner cube and fold mirrors for the robot TV cameras are located above the bearing assemblies. \label{fig6}}
\end{figure}

\begin{figure}
\epsscale{1}
\plotone{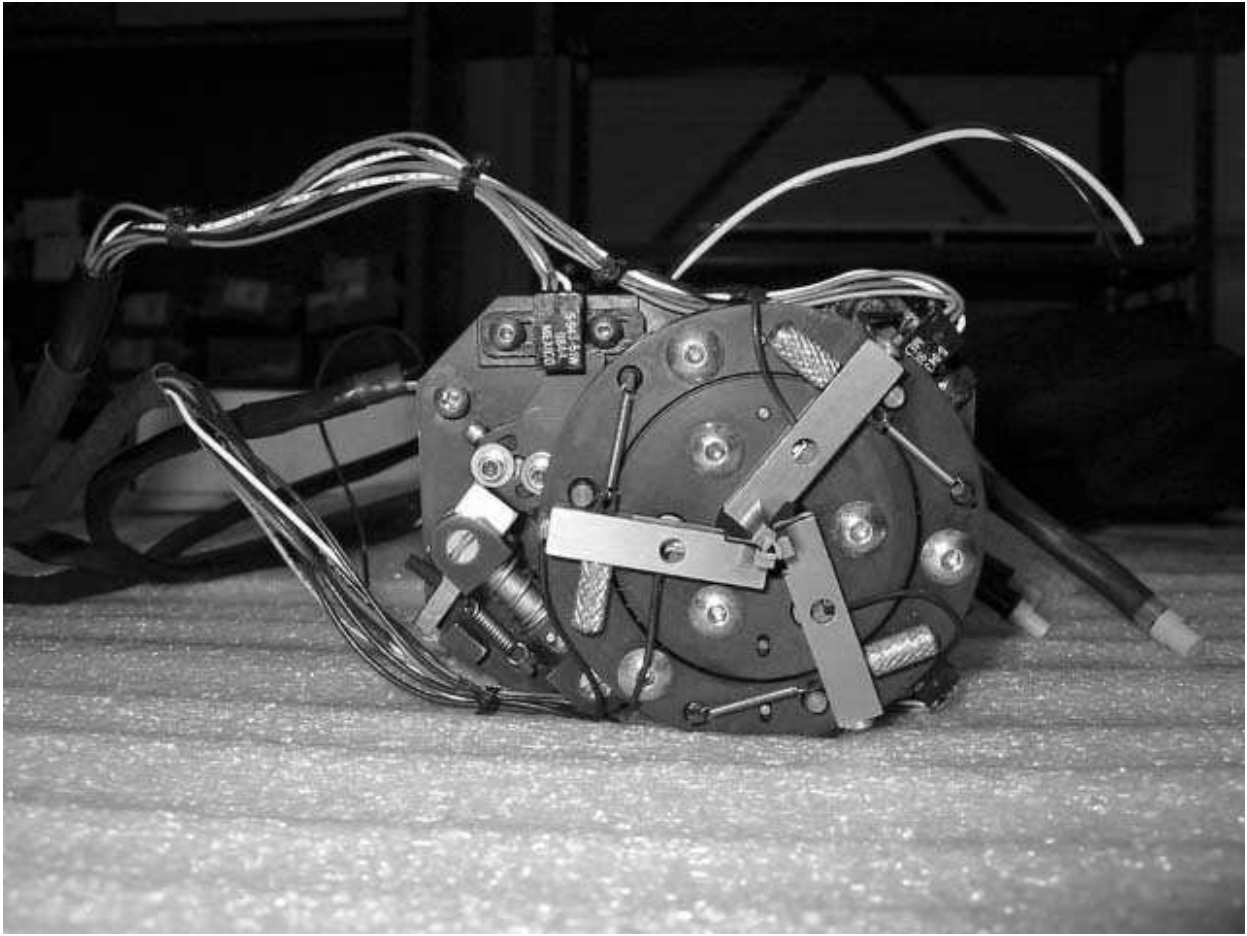}
\caption{Gripper assembly used to pick and place fiber buttons.  The three gripper jaws are actuated by a rotation of the outer ring to which the gripper fingers are attached, much like an iris mechanism.\label{fig7}}
\end{figure}

A pair of high-speed six-axis robots operating in tandem position Hectospec's
300 optical fibers. Each robot forms a three-axis Cartesian (XYZ) system that
carries a fiber gripper assembly. The Z-axis assembly (Figure \ref{fig6})
contains nested gimbals to allow placement of the fiber buttons perpendicular
to the curved focal surface, as well as the gripper mechanism. Figure
\ref{fig7} shows the gripper assembly removed from the Z axis assembly.  The Hectospec
gripper design is based on the earlier Hydra design \citep{Ba93}. Many
mechanical details including parts lists are found in the Hectospec Hardware
Reference Manual (http://cfa-www.harvard.edu/mmti/hectospec.html).

\subsection{Speed and Accuracy}

The timing budget for completing a Hectospec fiber repositioning consists of
eight steps: (1) coordinated XY$\Theta\Phi$ motion to the next button
position, (2) Z axis down, (3) gripper close, (4) Z axis up, (5) coordinated
XY$\Theta\Phi$ motion to the new button position, (6) Z axis down, (7) gripper
open, and (8) Z axis up. In total each fiber repositioning requires two
coordinated XY$\Theta\Phi$ motions, four Z axis motions, and two gripper
actuations. Table \ref{tblmovet} gives approximate times for these motions;
they total $\sim$1.8 s.

A combination of factors set Hectospec's internal positioning accuracy: (1)
the accuracy of the servo loop closure, (2) the calibration of the axis
encoders onto an orthogonal Cartesian coordinate system (includes axis home
reference repeatability), (3) button movement as the gripper jaws are
released, (4) the calibration of the fiber position within the button body,
(5) flexure of the focal surface from zenith to the observing position, and
(6) flexure in the guide probes over the course of an observation. Hectospec's
external positioning accuracy also includes contributions from astrometric
errors for the targets and guide stars and instantaneous tracking errors.
Table \ref{poserr} summarizes the measured and estimated magnitudes of the
internal positioning error sources. We set a goal of 0.025 mm for the total
internal positioning error; Table \ref{poserr} shows that the error
contributions sum to 0.015 mm in quadrature and 0.036 mm in a worst case
straight sum. The actual total internal positioning error lies in between
these estimates, or about 0.025 mm.

\subsection{Robot Safety Features}

The fiber positioning system has multiple levels of safety features built into
the hardware and the instrument control software. Our primary goal for these
safety features is to prevent collisions between the robots, collisions of the
robots with the focal surface or fibers on the focal surface, or high speed
collisions of the robots into their mechanical limits of travel. Additional
safety features act to minimize the resulting damage if a collision occurs. A
large number of lower level safety checks stop robot motion on the detection
of an error condition that could result in damage to some part of the
instrument.

\subsubsection{Collision Prevention}

The pair of robots start each move segment simultaneously and they wait until
both robots are finished before beginning the next move segment. (The eight
move segments are listed in Section 2.4.) Error conditions from either robot stop both robots.
During the paired moves the closest approach of the robots is 150 mm for the
entire fiber pick and place operation. Limit switches that detect approaches
of the robots closer than 150 mm or travel past the normal range of motion on
all axes provide the next level of safety. Energy absorbing bumpers on the X
and Y axes produce a controlled deceleration of the robots independent of the
electronic and software systems once the limit switches are passed. The most
difficult limit to detect is a downward movement of the Z axis, because the
appropriate limit position varies over the focal surface due to its curvature. We built
Z down limit switches into each of Hectospec's three gripper jaws; these
switches are activated if the jaws strike the focal surface or a fiber button.
The entire gripper mechanism is spring loaded; the gripper mechanism retracts
if the spring preload is exceeded.

\subsubsection{Lower Level Safety Features}

A key safety feature built into most servo controllers, including the Delta
Tau Programmable Multi Axis Controllers (PMACs) used for Hectospec, is
provided by the following error monitor. The following error is defined as the
instantaneous difference between the commanded and actual positions read back
from the encoder. The PMAC servo controller can be programmed to begin a
controlled deceleration when a preset following error is exceeded. Each of
Hectospec's servo axes is protected by a following error limit. The following
error protects against servo runaway if an encoder signal is lost and can
provide safe shutdown if an obstruction is encountered. The following error
provides no protection for moves shorter than the preset following error. An
additional level of protection is provided for Hectospec's powerful X and Y
axes: the signals from rotary encoders on the motor shaft and linear encoders
on the axes are continously compared with custom code in the PMACs and the
system is shut down if a small error is exceeded.

Other safety features include over-temperature sensing, a PMAC CPU watchdog
timer, amplifier fault detection, move time out protection, and a dropped
button sensor, all of which terminate motion upon error detection. Many levels
of error checking are also built into Hectospec's control software to prevent
setting a fiber button down on top of another button, to prevent striking a
fiber button or part of the positioner structure with the robots, or to move a
fiber button beyond its safe travel limits. 

\subsection{Robot Calibration Techniques}

The major portion of Hectospec's calibration was carried out with a custom
grid of dots etched into an 0.61 m diameter Astrosital disk by Max Levy Corp.
This calibration disk is the same diameter as Hectospec's focal surface and is
pinned and screwed into three support blocks in the guide probe tracks when in
use. The grid is illuminated with high frequency fluorescent lamps carried on
the robots, and the dot positions are recorded by the intensified cameras
carried in the robot gripper assemblies. After averaging for 5 to 10 seconds,
the noise in the dot position measurements is of order 2 $\mu$m. The dot
positions have an intrinsic positional accuracy better than 6 $\mu$m. The
positions of the dots recorded in the intensified robot cameras as the robots
are commanded to move to the nominal dot position allows us to transform the
XY encoder positions to an orthogonal Cartesian system. The largest errors
prior to calibration are $\sim$40 $\mu$m, a testament to the accuracy of
Hectospec's machining and to the quality of the axis rails. Most of this error
is due to: (1) small rotations of the robots as the gripper assembly moves
along its rails and (2) the position of the encoders at the guide rails, 150
mm above the position of the fiber button in the gripper jaws (and the
position of the grid dots during calibration). The spacing of the guide blocks
on the rails is also about 150 mm, so that deviations of the rail bed will map
$\sim$1:1 to deviations of the gripper jaws. This type of position sensing error is
commonly termed Abb${\rm \acute{e}}$ error.

The gimbal axes were calibrated using the robot intensified cameras and a fine
pattern of dots etched into a smaller grid placed on the focal surface. The Z
axis scale is accurately determined by a rotary encoder and a precision ball
screw driven through a 1:1 precision pulley system. A displacement measuring
laser interferometer was used to verify that this scale is accurate to better
than 1 $\mu$m RMS over the full range of travel. The ball screw accuracy is so
high that we removed an unnecessary LVDT intended for Z axis feedback. 
(A small slip of $\sim$1 $\mu$m upon each downward movement of the Z-axis is removed by detecting
a home sensor each time the Z-axis is raised.)
The
position of the focal surface relative to the robot axes was determined with a
Kaman noncontact displacement probe mounted to the gripper assembly. The
robots were commanded to move across the focal surface in a grid pattern, and
the position of the focal surface was determined at each point.

\subsection{Electronics}

The main design challenge for Hectospec's electronics was the distribution and
routing of cables associated with the 15 motor axes and their associated
encoders and limit switches. The weight and volume limits imposed by
Hectospec's Cassegrain mounting location precluded mounting the majority of
Hectospec's electronics onboard. The relatively large power supplies, servo
amplifiers, stepper motor drivers, cable distribution boxes, and signal and
power conditioning components are located in remote racks. A total of 20
cables with $\sim$460 total conductors run between the fiber positioner and these
racks, while 10 cables with $\sim$110 total conductors run between the bench
spectrograph and the electronics racks. To maintain high reliablity in the
signal path, heavy-duty Mil-C circular connectors are used where cables must
be disconnected to dismount the fiber positioner. The electronics rack
contains seven main electronics boxes: (1) an interface box that accepts the
Mil-C cables from the fiber positioner and distributes the signals to internal
rack cables, (2) an interface box that performs similar functions for the
bench spectrograph cables, (3) a signal conditioning and interface box for the
Delta Tau PMAC servo controllers, (4) a pair of servo electronics boxes that
contain the X,Y,Z,$\Theta$,$\Phi$ servo amplifiers as well as the gripper
stepper motor driver, (5) a stepper motor drive bay for the guider probe and
spectrograph stepper motors, (6) a power supply box, and (7) a power
conditioning and distribution module.

On the fiber positioner there are four electronics boxes: (1) the main
interface box that accepts most of the Mil-C cables and distributes signals
and power to internal cables, (2) a smaller interface box that accepts the
Mil-C cables for the X,Y,Z servo motors, (3) an electronics box with a CPU and
active circuits for on-board control functions, and (4) a small auxiliary box
with an over-illumination sensor to protect the intensified cameras as well as
power supplies for the fluorescent lamps used to illuminate the calibration
grids.

Surge and over-voltage protection is provided in each of the main interface
boxes at the fiber positioner, bench spectrograph, at the rack, as well as in
the rack-mounted power conditioning box. Further details can be found in the
Hectospec Hardware Reference Manual
(http://cfa-www.harvard.edu/mmti/hectospec.html).

\subsection{Guider Probes}

The guider probes each move one section of a trifurcated coherent fiber bundle
along an 86$^{\circ}$ arc just outside the focal surface plate that supports
the fiber probes at the MMT focus. \citet{fab98} describe the basic geometry
of the guider probes. The guider probes are actuated along a curved rail by a
stepper motor driving a pinion gear against a large-diameter fixed gear. The
guider probe has a brake mechanism, released by a solenoid, to hold the probe
in place after the guider probe is positioned with the telescope zenith pointing. The largest
challenge with this mechanism was designing a brake mechanism that would grip
firmly and that could be released with the limited force available from the
solenoid. Limited clearance from the structure precluded use of a large
solenoid, and we reworked this mechanism twice to reduce the friction in the
brake release mechanism. In future instruments pneumatic actuators might be a
good replacement for solenoids because the pneumatic actuators offer a better
power to weight ratio and do not dissipate as much heat.

\subsection{Instrument Control Computers}

The Hectospec fiber positioner is controlled by a Motorola 142 VME single
board computer running Linux. This computer is mounted in the same VME rack
that houses the two VME Delta Tau PMAC servo controllers. The {\bf hctserv}
server program running on the Motorola 142 accepts high and low level commands
from clients and issues the appropriate motion control commands to the PMACs.
Normally, the fiber positioner is operated by sending a high level sequence
command to move the fibers from one configuration to another. The sequence
begins when the client program reads the current fiber configuration from the
{\bf hctserv} server which is stored in static
memory. The client computes the sequence of fiber pick and place operations
required to go from the current to the new configuration, and prepares a
sequence table of paired pick and place moves for the two robots. The sequence
table is sent to the {\bf hctserv} server, which checks that the sequence
table can execute without crossing fibers or colliding the robots. Each paired
move is successively loaded into the PMAC and executed. 

A rack mounted Intel-based PC running Linux, ``Snappy", contains three Data Translation
DT3155 frame grabber boards that capture images from the two robot TV guiders
and the three guider probes. Snappy also communicates with the fiber positioner's internal electronic box
through an RS-232 interface to control the gain of the intensified TV guider
cameras, to control internal lamps, and to monitor motor temperatures. A
guider server running on a remote computer acquires guide frames from Snappy
and calculates guide corrections for the telescope.

\section{Hectospec Spectrograph}

\subsection{Mechanical Design}

\citet{fab98} describe the mechanical layout of Hectospec (see also
\citet{fab94}) and \citet{fata98} describe the optics mounts. Here we describe
the dewar assembly and rotary shutter not previously discussed in the
literature.

\subsubsection{Dewar Design}

\begin{figure}
\epsscale{1}
\plotone{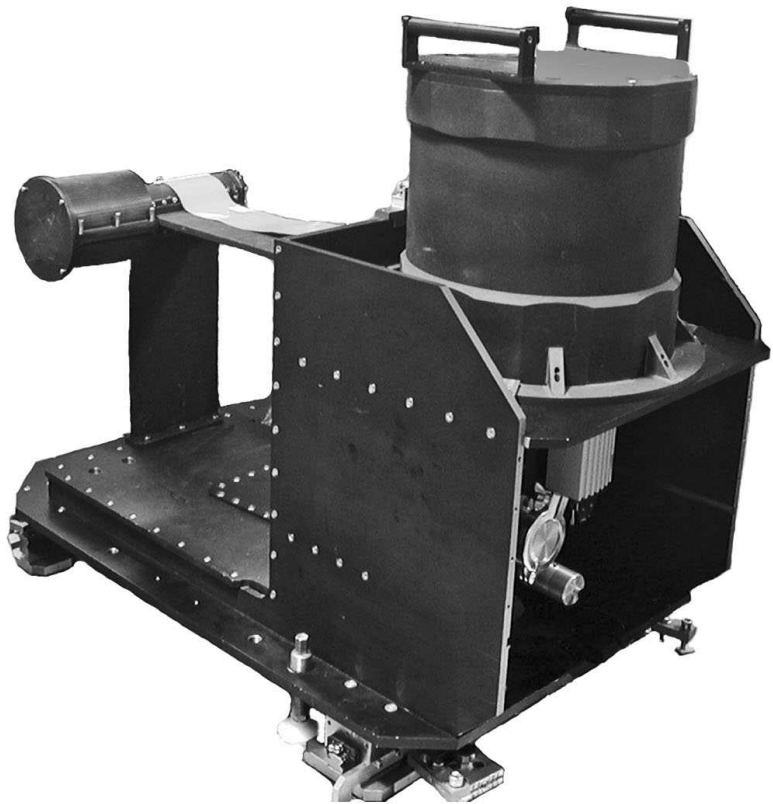}
\caption{The Hectospec dewar assembly.  The field flattener lens is obscured by a protective cover to the center left.  The evacuated tube surrounding the cold finger runs horizontally.  The LN$_2$ tank is prominent at the upper right.\label{fig8}}
\end{figure}

Hectospec uses an internal focus catadioptric camera (\citet{fab94}, \citet{fab98}).
Thus there is a premium upon mimimizing the footprint of the CCD dewar support
structure in the beam. The CCD dewar is an $\sim$100 mm diameter cylinder
supported on a thin vertical foot and by a thin-section horizontal evacuated
tube that surrounds the dewar's cold strap. The cold strap runs between the
dewar and a liquid N$_2$ cryostat mounted outside the optical beam running to and from
the on-axis camera mirror. The dewar window also serves as a field flattener
lens for the camera. Figure \ref{fig8} shows the dewar assembly. The dewar
assembly is mounted on a focus stage.

\subsubsection{Rotary Shutter}

\begin{figure}
\epsscale{.5}
\plotone{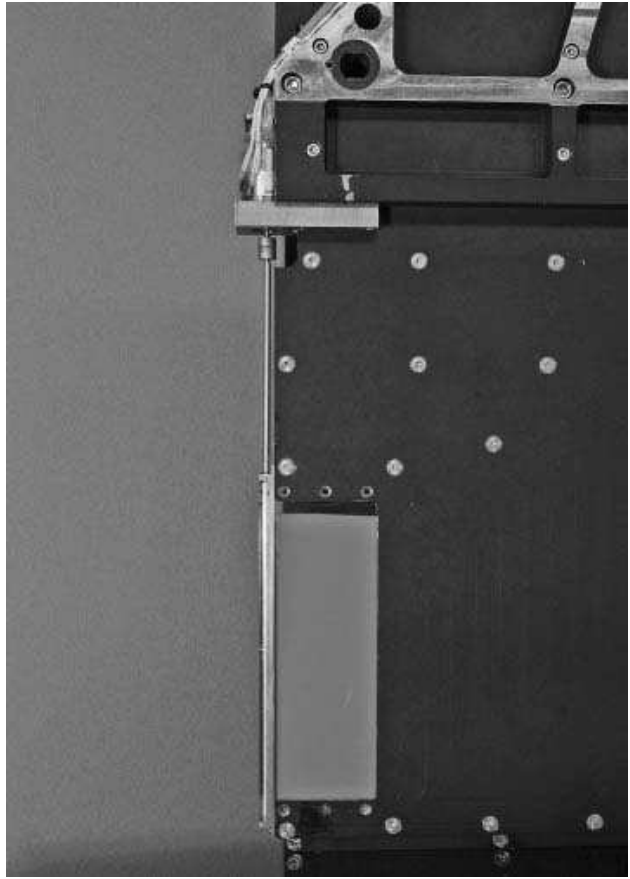}
\caption{The rotary shutter assembly mounted on the fiber shoe.  The small stepper motor used to drive the shutter is visible at the top center.  This motor is coupled to the rotary shutter mechanism with a long shaft.  The shutter itself is located at the lower left.\label{fig9}}
\end{figure}

At the spectrograph entrance ``slit", the optical fibers are arranged in 
two parallel columns spaced $\sim$1.6 mm
apart. The fibers in each column are spaced on 0.96 mm centers, but the two
columns are offset by 0.48 mm, giving a final effective fiber to fiber spacing
of 0.48 mm. The width of the structure holding the fibers, the ``fiber shoe",
has been minimized to reduce the obscuration of the beam returning past the
fibers from the on-axis collimator mirror. A narrow rotary shutter assembly is
mounted on the fiber shoe in front of the optical fibers. The rotating part of
the shutter is an $\sim$150 mm long slotted cylinder. The shutter is toggled
between open and closed with 90$^{\circ}$ rotation actuated with a stepper
motor. The rotary shutter assembly mounted on the fiber shoe is shown in
Figure \ref{fig9}. 

\subsection{CCDs and Array Electronics}

Hectospec uses two E2V model 42-90 4608$X$2048 CCDs with 13.5 $\mu$m pixels,
arranged in a 4608$X$4096 array. The long axis of the CCDs are parallel to dispersion.
Hectospec's CCDs have superb cosmetic
quality, and the entire two CCD array has only a single bad column. At a
readout rate of 100,000 pixels s$^{-1}$, the readout noise is 2.8 e$^{-}$
RMS. At the operating temperature of -120 $^{\circ}$C, the dark current is
very low: 1 e$^{-}$ in 900 s of integration. The electronics have a gain of 1
e$^{-}$ ADU$^{-1}$, matching to a few percent between the four amplifiers.
The count rate due to background radiation is 1 event s$^{-1}$ integrated over
the region occupied by the fiber spectra on the two CCD array. 

Hectospec shares its array controller electronics design with the other MMT f/5
optical and infrared instruments including Hectochelle, Megacam, and SWIRC.
The design of these electronics was driven by the requirements of the 72
channel Megacam version \citep{geary00}. Hectospec's two CCDs are read out
through four amplifiers in 46 sec.

\subsection{Optics}

\subsubsection{Image Quality}

\citet{fab98} and \citet{fab94} describe Hectospec's optics; here we describe
the optical performance of the as-built spectrograph. Hectospec's optics have
a reduction of 3.45, with an anamorphic demagnification (in the spectral
direction) of 1.06 with the 270 groove mm$^{-1}$ grating. With perfect optics,
the image of the 250 $\mu$m fiber should be an ellipse 5.4 by 5.1 pixels
across. The RMS image diameters produced by the real optics, when illuminated
by a point source emitting into an f/5.3 cone, are 1.3 to 1.8
pixels. Optical ray traces with ZEMAX predict that the azimuthally averaged
two dimensional FWHM of the fiber image should be 4.8 to 5.0 pixels, which
agrees with the observed azimuthally averaged FWHM. The tails of the
observed images are broader than the ZEMAX prediction: the observed 95\%
encircled energy radius is 4.2 pixels compared with the predicted 3.5 pixels.
We attribute this difference to small amounts of light emerging from the
fibers at focal ratios faster than the modeled f/5.3; this light is imaged in
the wings of the fiber image. We have a pupil mask at the grating to reject
most of this focal ratio degraded light, but the mask is somewhat oversized to
account for pupil rotation that depends on field angle. 

We discovered one general feature of the HIRES style of camera \citep{hires93}
used in Hectospec: a ghost pupil is formed between the focal surface and the
field flattener element by light reflected from the CCD and then back again
from the front surface of the field flattener. The intensity of this ghost
pupil is reduced by a factor of $\sim$10$^{-3}$ relative to the main image and
the ghost pupil area is nearly as large as the detector area. We noticed the
ghost pupil because we operated the spectrograph briefly with a damaged
antireflection coating on the front surface of the field flattener. The
reflectivity of the damaged antireflection coating was about an order of
magnitude larger than normal, increasing the intensity of the ghost pupil by a
corresponding factor. Once we replaced the field flattener with a undamaged
spare, the ghost pupil faded into obscurity, but it may be wise to keep an eye
on this issue in designs of this family of optics. 

\subsubsection{Optical Coatings}

The initial optical coatings for the Hectospec spectrograph were developed for
the FAST spectrograph \citep{fast} at Whipple Observatory's 1.5 m Tillinghast
Telescope. FAST's high throughput Sol-gel antireflection coatings and
UV-enhanced overcoated silver reflection coatings gave many years of service
and we expected the same performance in Hectospec. Hectospec's Sol-gel
antireflection coatings have been trouble free, but the first protected silver
coatings on the Hectospec mirrors were noticeably discolored by late 2002,
approximately three years after their coating. Reflectivity measurements
indicated low reflectivity between 5000 and 6000 {\rm \AA}. We decided to
strip the original coatings and to recoat with the Lawrence Livermore National Laboratory (LLNL) durable silver
reflective coating (US Patent 6,078,425). The LLNL coating has slightly lower peak
reflectivity than the original (fresh) coating, but offers much better
reflectivity below 4000 {\rm \AA}.

\subsection{Gratings}

Hectospec's 259 mm collimated beam diameter requires large custom ruled
gratings. Hectospec currently has two gratings: a 270 groove mm$^{-1}$ grating
blazed at 5000 {\rm \AA} and a 600 groove mm$^{-1}$ grating blazed at 6000
{\rm \AA}. In both cases we specified a 5000 {\rm \AA} blaze, but the blaze
angle has proven to be the most difficult parameter to control in ruling these
large gratings. The 6000 {\rm \AA} blaze for the 600 groove mm$^{-1}$ grating
was the closest to specification achieved in four
attempts. The 270 groove mm$^{-1}$ grating provides a dispersion of 1.2 {\rm
\AA} pixel$^{-1}$ and a resolution of 6.2 {\rm \AA} FWHM. The 600 groove
mm$^{-1}$ grating provides a dispersion of 0.5 {\rm \AA} pixel$^{-1}$ and a
resolution of 2.6 {\rm \AA} FWHM. Both gratings have a peak absolute
efficiency of 72\%. Higher ruling density gratings might be best provided by
assembling grating mosaics.

\section{Hectospec Optical Fibers}

\subsection{Design of Fiber Run from Positioner to Spectrograph}

The design of the optical fiber run between the fiber positioner and bench
spectrograph turned out to be a much larger challenge than we had imagined
during the Hectospec's conceptual design phase. Optical fibers offer a unique
means of mapping a spectrograph slit efficiently onto a large number of
objects distributed over a very large field of view. However, fibers must be
used carefully if they are not to severely compromise the sensitivity of the
spectrograph by degrading the focal ratio of the light incident on the fiber.
Degradation of the focal ratio in a fiber results in a loss of light
in the spectrograph optics unless the spectrograph optics are significantly
oversized. Oversizing the optics is expensive in a spectrograph as large as
Hectospec's, and forming good images from the light scattered into faster
focal ratios makes a difficult optical design problem harder. Focal ratio
degradation arises from mechanical stress of the fibers. Focal ratio
degradation will typically cause time variable throughput losses as the stress
experienced by the fibers changes as the telescope pointing direction changes.
These throughput variations are doubly troublesome because good sky
subtraction relies on an accurate fiber to fiber throughput calibration.

\begin{figure}
\epsscale{1}
\plotone{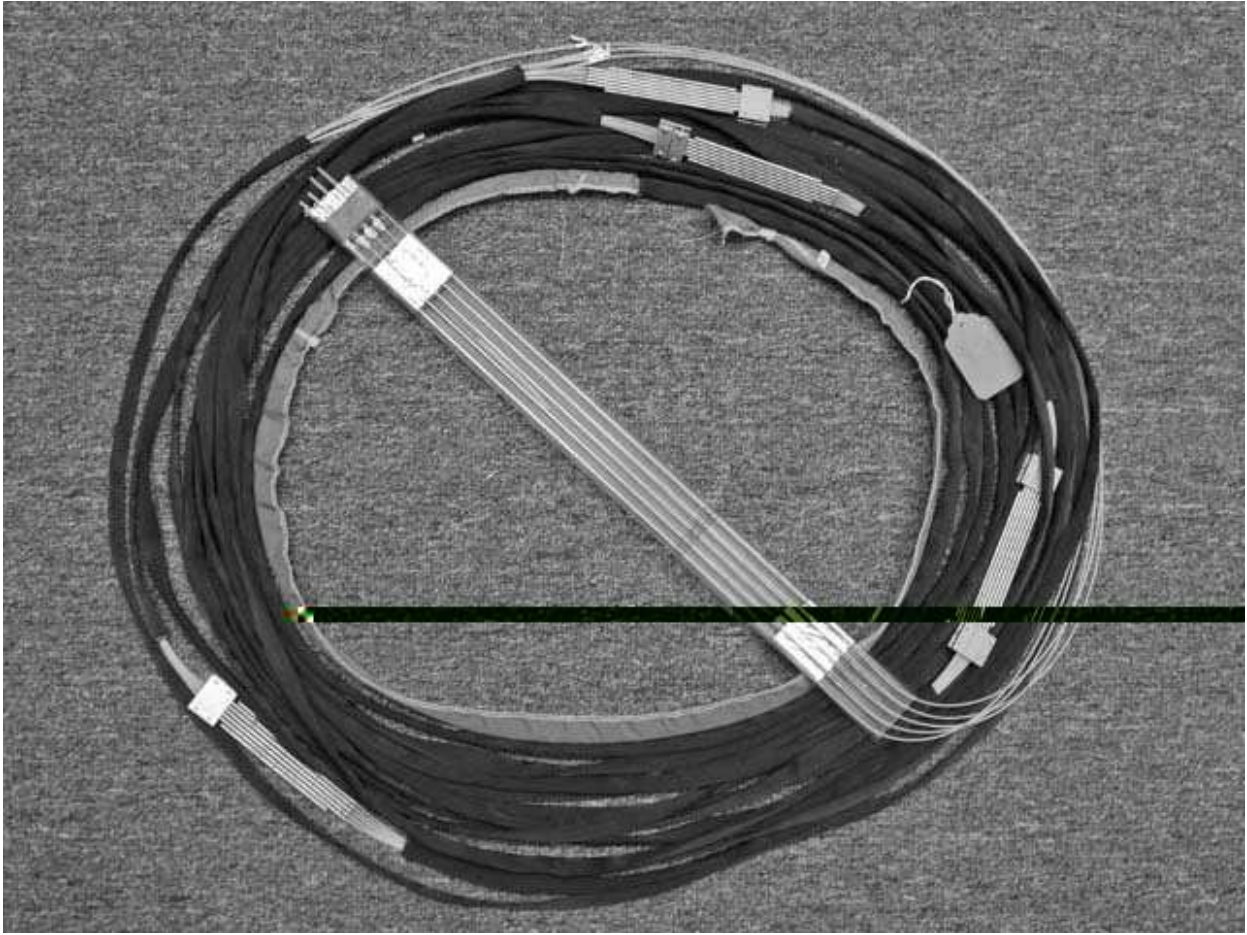}
\caption{Coiled 26 m fiber run.  The four thermal break assemblies are located at one, four, and eight o'clock.  The fiber probe assemblies, protected by plastic tubes, are laid across the center running from the lower left to the upper right.\label{fig10}}
\end{figure}

\begin{figure}
\epsscale{1}
\plotone{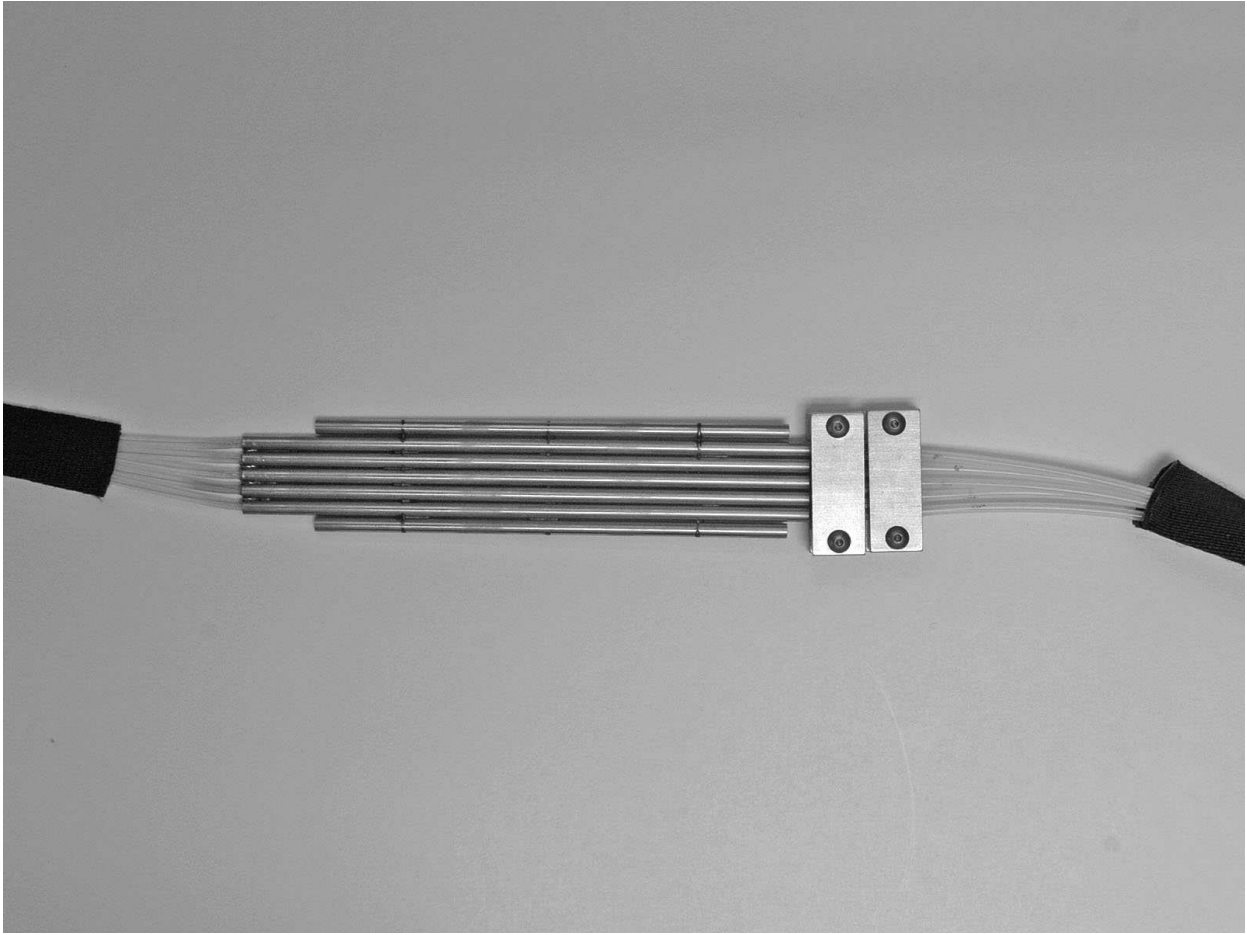}
\caption{One of four fiber thermal breaks in the fiber run.  A shipping clamp restraining the separation of the teflon tubes from the stainless thermal break tubes is in place at the right.\label{fig11}}
\end{figure}

We were aware from the time that we began work on Hectospec that the mounting
details of the fibers in the fiber probes at the fiber positioner end and in
the fiber slit at the spectrograph end of the fibers are important in
controlling focal ratio degradation, but our test program demonstrated the
importance of what happens in between. We engineered thermal breaks in the
teflon tubing to address the thermal expansion mismatch between the fused
silica fiber and the teflon tubes that protect the fibers. \citet{fab98}
describe the basic design elements that we used to control focal ratio
degradation in the fiber run. The other important design driver for the fiber
run is that it must allow easy mounting and dismounting of the Hectospec on
the telescope. Figure \ref{fig10} shows the design of the fiber run and the
position of the thermal breaks that allow for the differential thermal
expansion of the fibers and their protective teflon tubes. Figure \ref{fig11}
is a closeup of one of the four thermal breaks in the fiber run.

\subsection{Laboratory Tests of the Optical Fibers}

\begin{figure}
\epsscale{1}
\plotone{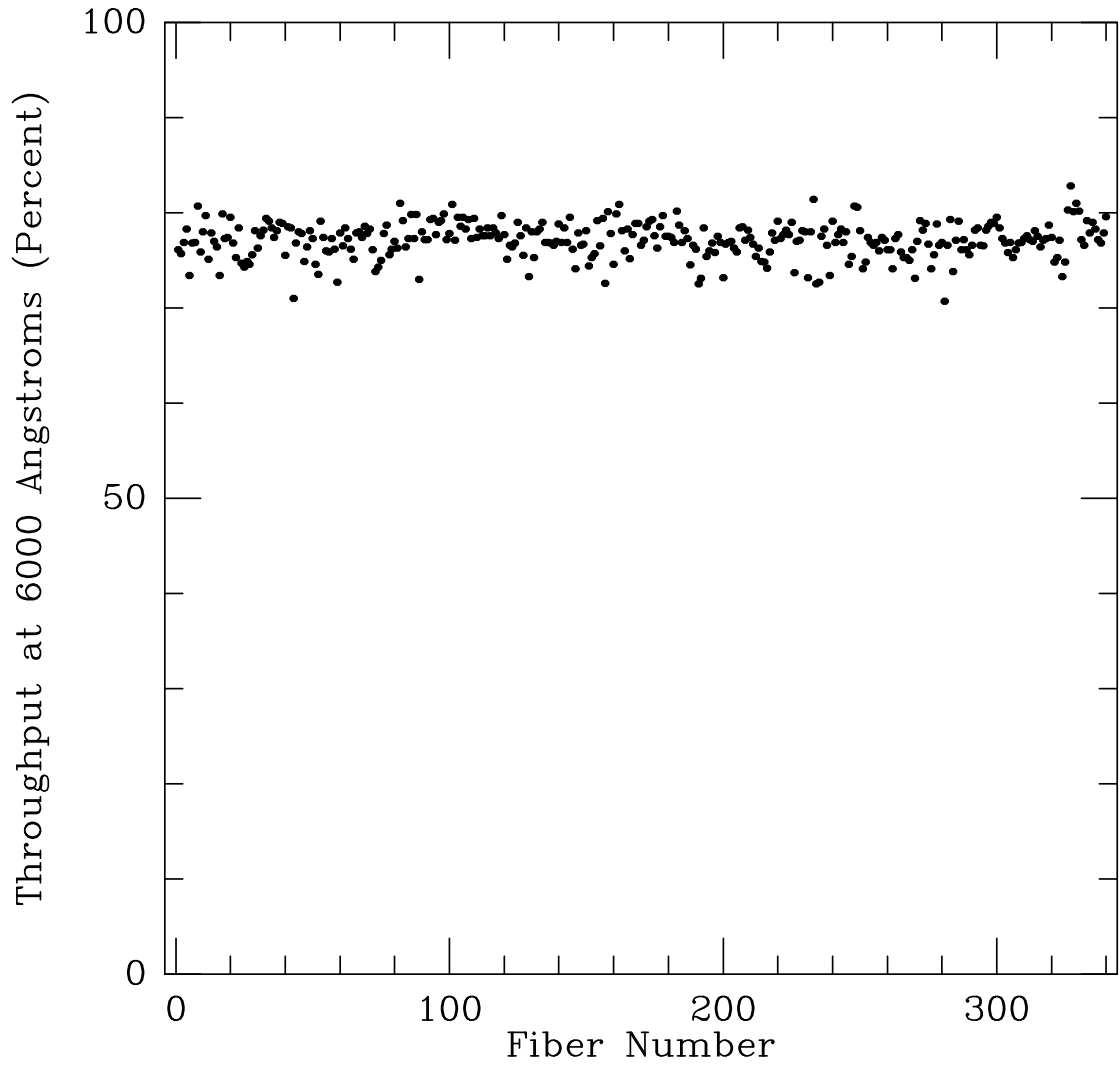}
\caption{Fiber throughput at 6000 {\rm \AA} as measured in the laboratory.\label{fig12}}
\end{figure}

Our specifications called for a total fiber throughput of 75\% at 6000 {\rm
\AA} into a focal ratio of f/5.3, the final focal ratio of the spectroscopic
wide field corrector and the design focal ratio of the Hectospec bench
spectrograph. At 6000 {\rm \AA} the internal absorption of Hectospec's 26 m long fibers is 6\%,
the Fresnel reflection loss at the uncoated BSM51Y prisms at the fiber input
end is 5.3\%, and the Fresnel loss at the bare fiber output end is 3.5\%. The
total throughput accounting for only these three fiber losses is 86\%,
requiring that 87\% of the light entering the fibers at f/5.3 exit within an
f/5.3 cone if the total throughput specification is to be met. As Figure
\ref{fig12} shows, we met our goal for total fiber throughput with an average
throughput of 77\% and a standard deviation of 2\%.

\section{Observing with Hectospec}

\subsection{Observation Planning Software}

\citet{roll98} describe the basic algorithms used in Hectospec observation
planning to match fiber probes to targets. Currently, each Hectospec observer
is responsible for planning the fiber configurations to be used at the telescope.
The observer begins by assembling a catalog containing the desired targets as
well as potential guide stars in the field. Observers may priority rank
targets in the catalog and assign a minimum and maximum number of fibers to
each rank. The guide stars and the targets must have excellent relative
astrometry because the guide stars establish the field alignment. The observer
uses the {\bf xfitfibs} program to create instrument configuration files from
the input catalog. {\bf xfitfibs} combines the observation planning software
with a graphical user interface (GUI). The observer may use {\bf xfitfibs} to
plan multiple fiber configurations with multiple field centers and to rank
these centers in priority.

In addition to matching fibers to targets, {\bf xfitfibs} selects guide stars,
and assigns fibers for the measurement of the sky background. The user can
drag field centers with a mouse or enter new field centers into a table and
view the effects on a graphical display. The display indicates guide stars
within range of the guide probes as well as those accessible by changing the
instrument rotator angle. The {\bf xfitfibs} software derives guide star
magnitudes from either the 2MASS or GSC2 catalogs, removes stars outside the
specified magnitude range, and aditionally removes non-stellar objects using
GSC2 classifications or by SExtractor classification from DSS images. Stars
with close companions or brighter neighbors in the field of the guide probes
are also excluded. Selection of bright guide stars (R$<$16) minimizes setup
time with the intensified robot and guide cameras and allows operation in the
full moon.

\subsection{Queue Operation}

Hectospec is operated in queue mode to use valuable telescope time efficiently and to distribute the time lost due
to weather, telescope and instrument malfunctions evenly. The goal of the queue scheduling is to distribute the
available productive observing time proportionally to the nights scheduled by
the Time Allocation Committees. The MMT is scheduled in trimesters, and the
queue scheduling is for a single trimester at a time. The queue schedule must
be continually updated through an observing run in response to the weather
conditions and telescope/instrument performance. The observers provide the
queue manager with completed fiber configurations, which include the sequence
of observations, the fiber layout, and valid guide stars. These layouts are
checked and then scheduled as time permits. The fiber positioner is operated
by experienced robot operators on behalf of the queue. The robot operator also
oversees the acquisition of calibration data in the late afternoon and early
evening including HeNeAr wavelength calibrations, incandescent lamps for
flatfielding, and typically twilight sky flatfield exposures. During the
night, observers assigned telescope time operate the spectrograph and evaluate
the data quality on behalf of the queue. Observers thereby gain firsthand
experience with Hectospec's operation and capabilities, avoiding some of the
communication difficulties that others have reported in the operation of
observing queues isolated from the scientist who will ultimately use the data.

\subsection{User Interfaces}   

\subsubsection{Hobserve}

\begin{figure}
\epsscale{1}
\plotone{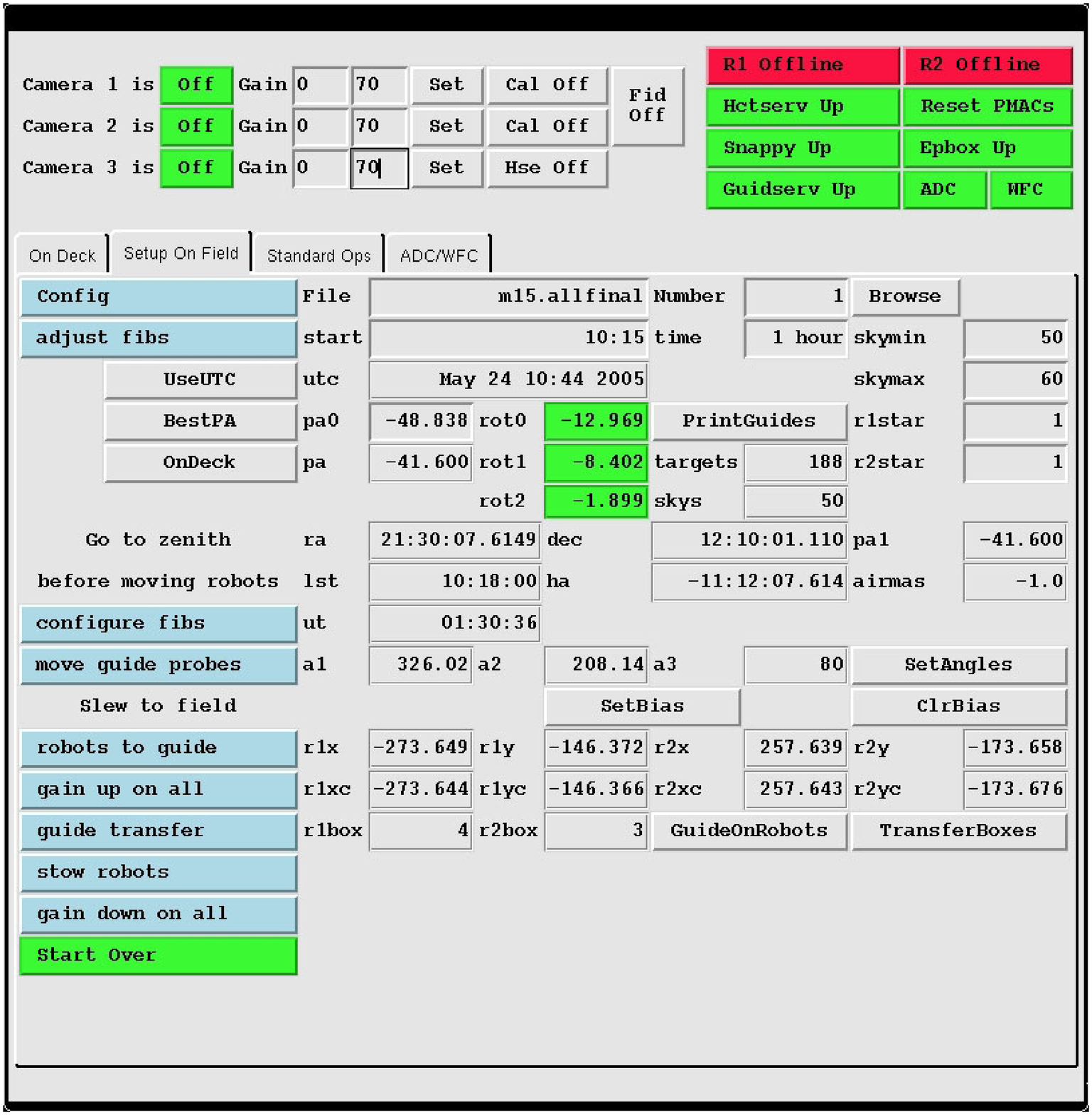}
\caption{The main screen of the Hobserve GUI. {\bf Hobserve} is a tcl/tk script that provides a step by
step procedural interface that guides the robot operator through the procedure of
calculating the appropriate fiber configuration for the chosen observing time,
calculating the sequence of fiber moves required to attain that configuration,
issuing the move commands to the robots, and setting up the guiding on the
field with the preselected guide stars.\label{fig13}}
\end{figure}

\begin{figure}
\epsscale{1}
\plotone{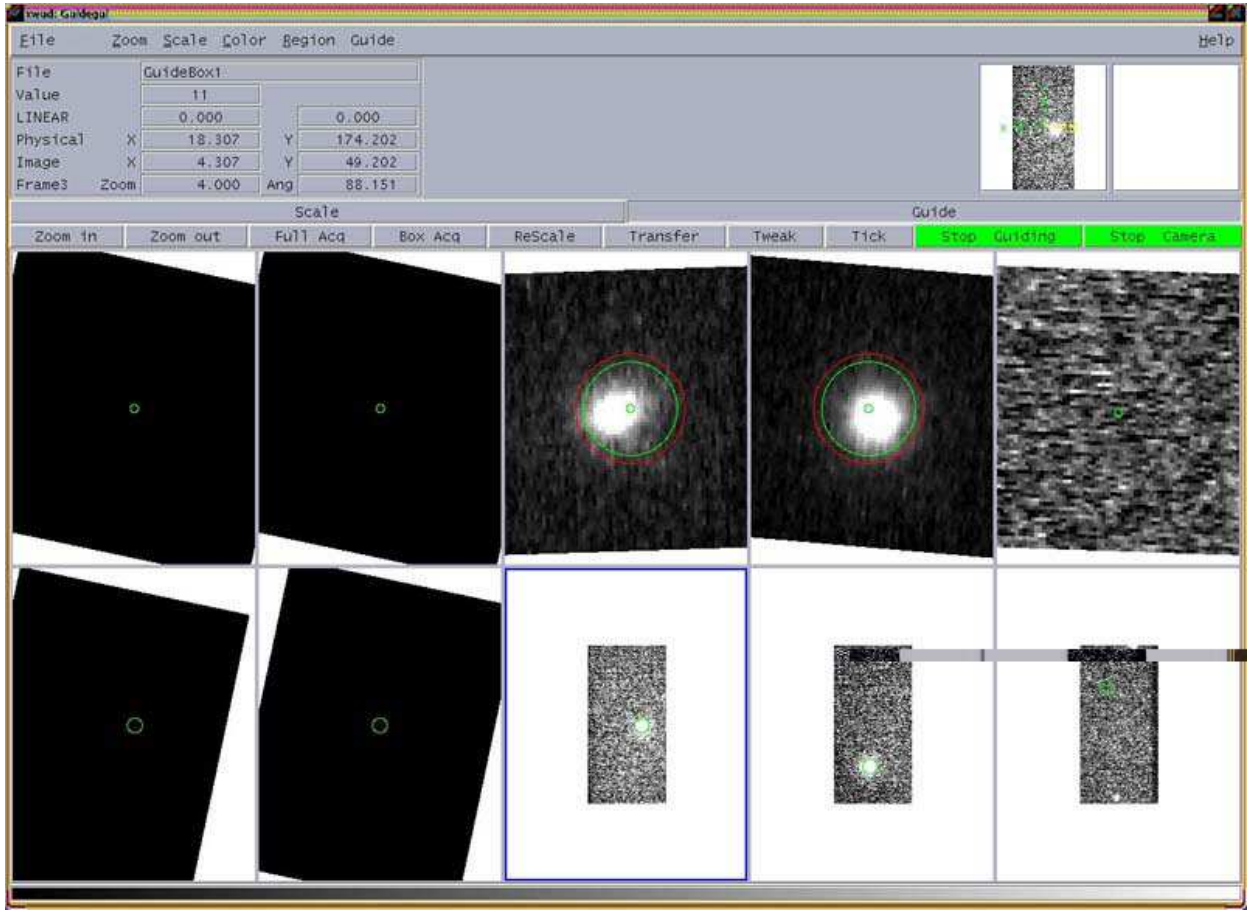}
\caption{The guidegui display showing the guide star images from the guider probes and the reference positions.  The robots are stowed during
normal observations, so no guide stars are visible in the leftmost windows.\label{fig14}}
\end{figure}

{\bf Hobserve} (Figure \ref{fig13}) is a Tcl/Tk script that provides a step by
step procedural interface that guides the robot operator through the procedure of
calculating the appropriate fiber configuration for the chosen observing time,
calculating the sequence of fiber moves required to attain that configuration,
issuing the move commands to the robots, and setting up the guiding on the
field with the preselected guide stars. A separate graphical user interface (GUI) ({\bf guidegui} (Figure
\ref{fig14}) displays the guide star images superposed on the guider target
positions. The {\bf Hobserve} script coordinates the actions of several
servers and calls other high level shell scripts to accomplish its functions. 

\subsubsection{SPICE}

\begin{figure}
\scalebox{.75}{\includegraphics[angle=-90]{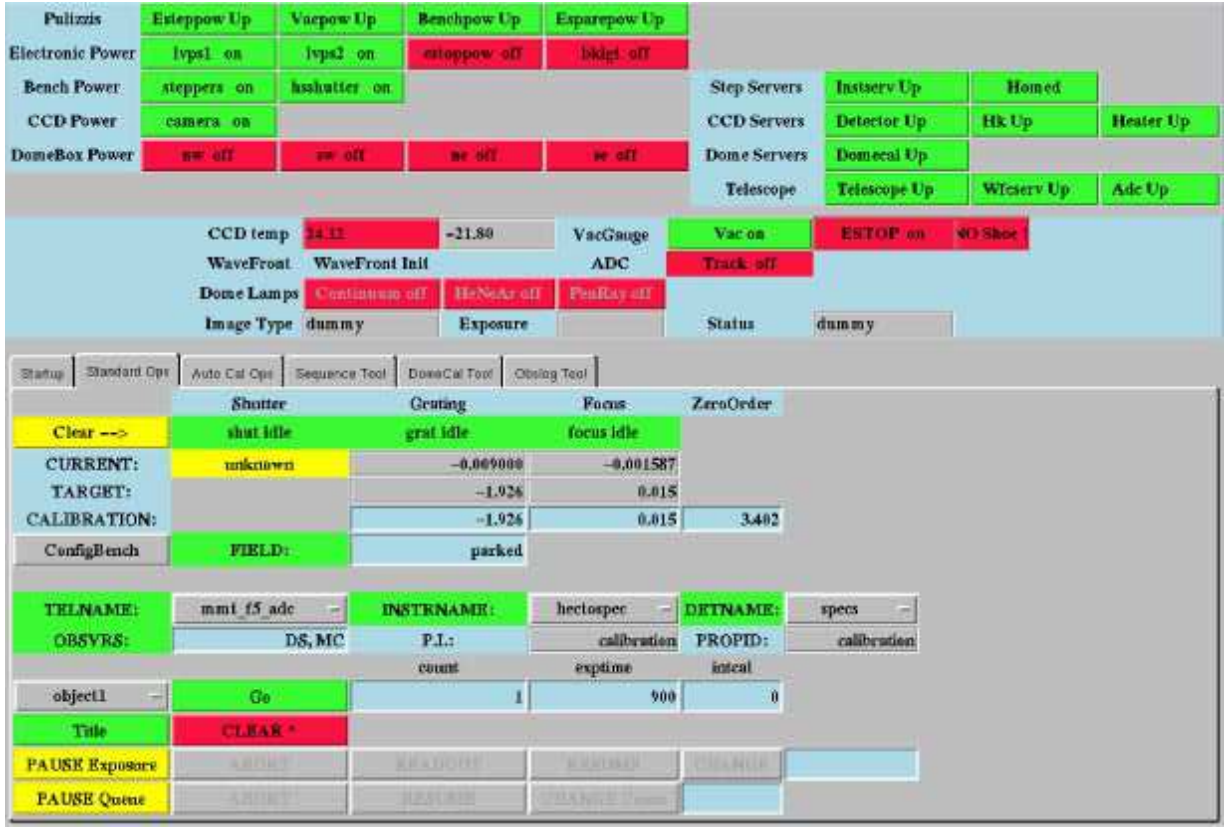}}
\caption{The {\bf SPICE} program provides a GUI interface to the
spectrograph and camera controls and sequences the operations for various
types of exposures. The GUI consists of two fixed displays at the top of the
window along with several tab selectable displays that appear at the bottom of
the window.\label{fig15}}
\end{figure}

The {\bf SPICE} Tcl/Tk program (Figure \ref{fig15}) provides a GUI interface to the
spectrograph and camera controls and sequences the operations for various
types of exposures. The GUI consists of two fixed displays at the top of the
window along with several tab selectable displays that appear at the bottom of
the window. The upper fixed display indicates the status of the system power
controls and of the software servers that control hardware components. The
text message within the status buttons and the color of the status buttons
change together to indicate the instrument state. Toggling these buttons
toggles the state of the power or software server. The lower fixed display
provides housekeeping information as well as exposure status. The housekeeping
items include the dewar temperature, the status of the calibration lamps, the
tracking status of the ADC prisms, and the position of the wave front sensor
carriage. The exposure status information includes the current exposure count
within the requested number of exposures, the current exposure type (skyflat,
object, bias, dark, etc.), and the current exposure state (exposing, reading
out, etc.).

The observer can select one of four user-selectable tabbed displays: an
initialization tool, a standard operations tool, a calibration lamp tool, and
an observing log tool. The initialization tool controls the sequencing of
powering up and homing the three spectrograph axes (shutter, focus, and
grating stage), as well as bringing up the required software servers. 

The standard operations tool is the main tabbed display window that the
observer uses to acquire data and control the spectrograph. Within this window
the grating rotation, focus position, and shutter status are displayed and can
be controlled. The current fiber configuration file name is displayed, but the
fiber configuration is controlled from the {\bf Hobserve} interface. The
observer selects the desired exposure type from a pull-down menu (object,
dark, bias, domeflat, HeNeAr, focus, etc.) and specifies the desired exposure
time and number of exposures. Pressing ``Go" initiates the exposure sequence.
During the initialization of the exposure, the software checks for possible
problems including a mismatch between the requested and encoded grating
rotation and focus position, blocking by the wavefront sensor, illumination of
calibration lamps at inappropriate times, lack of guiding signals, and ADC
prism tracking turned off. The focus procedure automatically moves the grating
to the zero order position and takes a sequence of CCD charge-shifted
exposures at different focus settings. The grating is returned to its nominal
position at the end of the focus run.  Focus changes during a run appear to be
negligible.

The calibration lamp tool allows the observer to control the calibration lamps
that illuminate a screen mounted on the MMT building shutters. The calibration
lamps are mounted in four identical lamp boxes mounted in the MMT chamber.
Lamp status, lamp voltages, and lamps currents are displayed in a
comprehensive color-coded table.

The observing log tool provides a summary of each of the exposures in the
selected night's directory and a text box for the observer to annotate each
exposure. A PostScript formatted observing log of all the exposures (with observer comments) for the
current night is produced and archived.

\subsection{Observing Sequence}

At the beginning of the night or during an exposure on the previous field, the robot operator loads in the
next fiber configuration in the ``on-deck" window of {\bf hobserve}. The expected
start time and exposure duration of the observation are entered, and the fiber
positions are retweaked for the airmass and instrument rotator position at the
expected exposure midpoint. Depending on the nominal position angle entered
during observation planning, the rotator demand angles may exceed the
$\pm$90$^{\circ}$ soft limits.  Therefore, the position angle can be reset by the robot operator to minimize
the rotator demand angles. If guide stars are available in two of the three
guide star probes, the ``on-deck" configuration is ready to go. If two guide
stars are not available at this position angle, the position angle can be
manually adjusted to find minimum rotator demand angles consistent with
obtaining two guide stars. In the worst case, if two guide stars are not
available at acceptable rotator angles, another configuration can be chosen
without loss of observing time. 

We begin the reconfiguration process by slewing the telescope to the zenith.
The robots are capable of operating at any combination of zenith angle and
rotator angle, but we reconfigure the fibers with the telescope zenith
pointing to minimize the power dissipation and load on the drive components.
After the robots reposition the fibers, the guide probes are moved to correct
guide star positions. The telescope is then slewed to a bright star near the
field position, and the wave front sensor is deployed to correct the primary
support forces and telescope collimation. While the wave front sensor probe is
retracted, the telescope is slewed to the field position and the instrument
rotator is set to track the desired position angle. The robot guide cameras
are sent to the guide star positions and the telescope pointing and the
rotator angle are adjusted with autoguiding software until the guide stars
appear at the positions corresponding to a fiber held
in the gripper jaws. A pellicle beam splitter in the robot optical train
allows 50\% of the light to pass through to the guide probes. The positions of
the guide stars in the guide probes are captured and the guiding is then
transferred to the guide probes. This procedure directly ties the guiding to
the robot-defined coordinate system upon which the fibers are positioned. As
soon as the robots are parked an observation can begin. 

\section{Hectospec Performance at the MMT}

\subsection{Throughput}

\begin{figure}
\epsscale{1}
\plotone{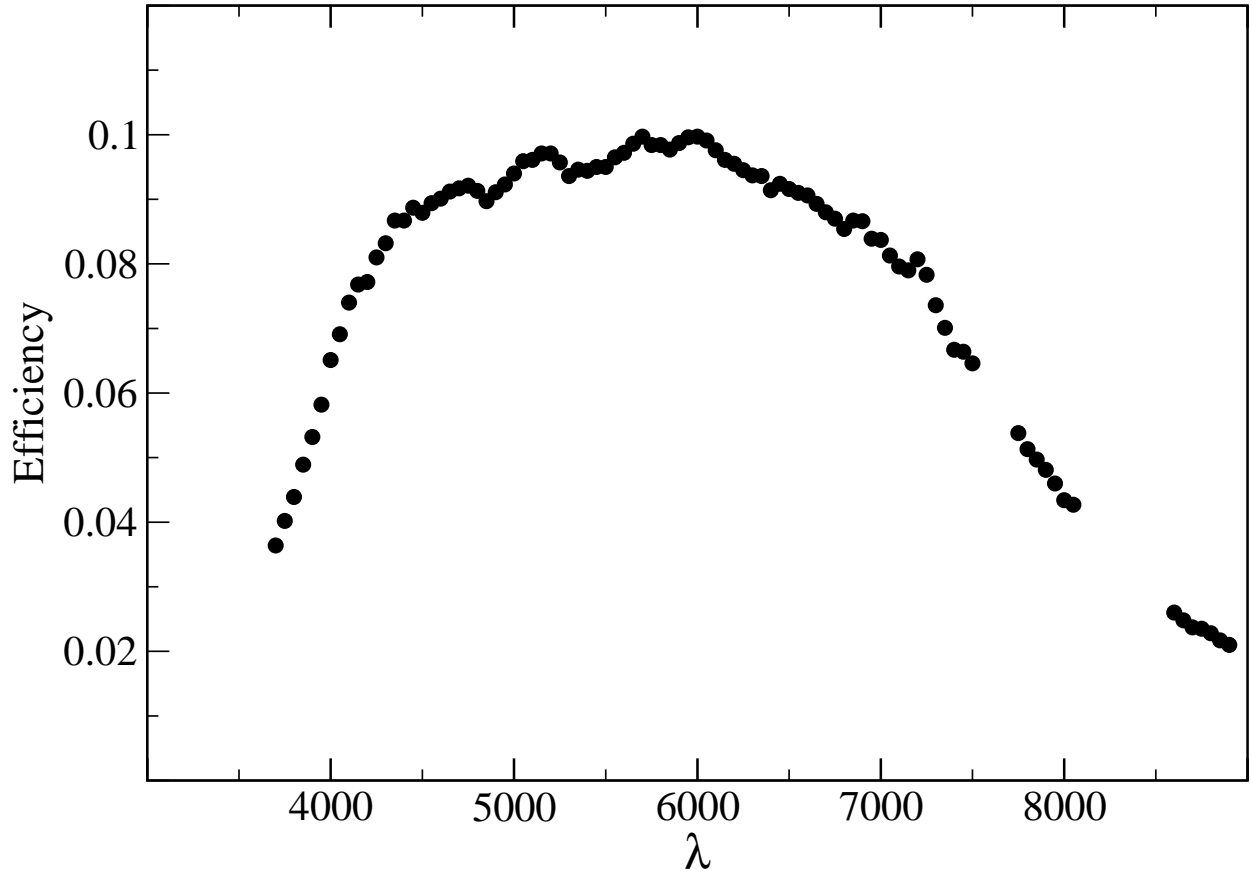}
\caption{The measured system throughput as a function of wavelength in 1$^{\prime\prime}$ seeing. Aperture losses and losses in the telescope optics are included.\label{fig16}}
\end{figure}

We measured the throughput of Hectospec with the spectrophotometric standard
star BD+284211, stepping the star across one of Hectospec's fibers to
determine the point of maximum throughput to eliminate light loss due to
astrometric errors. With a direct spectrograph, the entrance aperture is
normally opened as wide as possible to separate the measurement of
spectrograph throughput from aperture losses. With fibers, the entrance
aperture is fixed and we measure total throughput including aperture losses.
Figure \ref{fig16} shows the derived throughput measurement for BD+284211 in
1$^{\prime\prime}$ FWHM seeing. We have separately calculated the aperture
losses for Hectospec's 1.5$^{\prime\prime}$ diameter fibers as a function of
seeing using images obtained with Megacam; in 1$^{\prime\prime}$ FWHM seeing
we find that a 1.5$^{\prime\prime}$ diameter contains 59\% of the light,
giving an aperture correction of 1.7. Referring to Figure \ref{fig16} and
applying this correction, the system throughput peaks at $\sim$17\% between
5000 and 6000 {\rm \AA}. Our prediction for the total system throughput peaked
at 20\% in the same wavelength region, in reasonable agreement with the
measurement. 

\subsection{Thermal Flexure of Bench Spectrograph}

The room that encloses the Hectospec and Hectochelle bench spectrographs is
one level above the telescope chamber floor, and shares a common wall with the
telescope chamber. The spectrograph room is not well insulated from the
telescope chamber due to several doors and perforations. As a result, the
temperature in the spectrograph room fluctuates considerably more than we had
planned. The nightly temperature swings lead to temperature gradients in the
optical bench and optical mounts that lead to thermally-induced image motion
with a maximum excursion of one pixel in both the spatial and spectral
directions during the course of a night. Larger shifts of a few pixels occur
over the course of an observing run. As discussed in Section 7.2, the shifts
during a night are easily corrected by tracing the positions of the fiber
spectra and by tracking the positions of night sky emission lines. During
August 2005, we will replace the coated black cloth currently enclosing the
bench spectrographs to prevent light leaks with light-tight insulated panels.
This replacement will remove a potential fire hazard from the light-tight but
flammable cloth and improve the thermal insulation in the spectrograph room.

\subsection{Gravitational and Rotator-Induced Flexure of Fiber Positioner}

Our finite element models of Hectospec predicted that the Hectospec's focal
plane assembly (onto which the fibers are attached) would shift by $\sim$50 $\mu$m
relative to the robot's gripper assembly as the telescope pointing changes
from zenith to an elevation angle of 30$^{\circ}$. This flexure is relevant
because we use the robots to establish the correct position of the guide stars
in the guider probes. We took advantage of a cloudy night to measure this
flexure by placing the fibers on the focal surface at the zenith in the usual
fashion, tipping the telescope in elevation, and measuring the position of the
backlit fibers. The average deflections that we measured moving from an
elevation angle of 90$^{\circ}$ to an elevation angle of 30$^{\circ}$ are 40
$\mu$m in one robot and 42 $\mu$m in the other. We also noticed that rotating
the instrument rotator produces repeatable deflections of up to 25 $\mu$m if
the elevation angle is held constant. Rotator-induced deflections were
anticipated, and set stringent requirements on the accuracy of the rotator
bearing. 

\subsection{Observing Overhead}

The elapsed time between completing an exposure on one field and beginning
another is ~1040 s. Table \ref{overhead} lists the items that contribute to
the observing overhead. The observing overhead will drop slightly with
increased automation of the wave front sensing procedure, but is unlikely to
drop much below 900 s.

\subsection{Hectospec Reliability}

Hectospec has proven to be reliable at the telescope despite its complexity.
We have lost $\sim$4 nights of observing time in Hectospec's first year of
operation due to instrument malfunctions. The most serious incident occurred
in July 2004. A backlash removal spring in the gripper mechanism failed, and
the gripper misplaced a button. During the diagnosis and recovery process, a
series of low level robot commands were issued and the robots were
inadvertantly parked with a fiber caught in the gripper jaws. This event
resulted in the spectacular destruction of the captured optical fiber. The
damaged fiber was replaced with a spare fiber from one of the five spare
groups of fibers, and the antibacklash springs in both grippers were replaced
with longer springs of the same spring constant. Our analysis revealed that
the original spring was repeatedly overstressed in normal operation. We have
also revised our procedures to require visual inspection of the Hectospec
focal plane and fiber positions before executing low level robot commands.

Two of the other Hectospec malfunctions involved the guide probes. During the
first incident, the guider probe braking mechanism on one of the probes stuck
open, causing the guide probes to fall out of position as the telescope was
slewed away from the zenith. The problem was traced to excessive friction on
the brake guide shafts which was removed by polishing the shafts on all three
guider probes. The second incident was caused by the guider brake failing to
release on one of the guider probes. This problem was traced to a loose
connector, which was possibly not fully seated following the brake shaft
rework.

\section{\label{pipeline} Hectospec Data Analysis Pipeline}

\subsection{Introduction}

The $\sim$5000 spectra produced nightly by the Hectospec are reduced using a
semi-automated pipeline; typically this reduction is completed by the
afternoon following the observations.  The reduction yields a catalog of redshifts, errors,
signal to noise estimates and crude spectral types in time to modify the next
nights observing plans, if necessary.
\citet{mink} give an overview of the entire data flow, from forming the
observing catalog to archiving and releasing the final reduced spectra. Here
we describe those steps taken to reduce the raw CCD frames into a set of
wavelength calibrated spectra with redshifts.
The system rests on the IRAF spectral reduction packages \citep{valdes92} and
the IRAF task {\bf dofibers} \citep{valdes95}. The process is controlled by a
set of custom IRAF CL scripts, similar to those which control the reduction of
data from the FAST spectrograph \citep{tokarz97}.

\subsection{Wavelength Calibration}

Accurate wavelength calibrations are crucial to the performance of fiber
spectrographs on large aperture telescopes. Small wavelength
calibration errors cause artifacts in the subtraction of the near infra-red OH
night sky lines which mask or mimic the appearance of H$\alpha$ emission at $z\ge0.2$. 
In a one hour exposure Hectospec is routinely able to measure spectra for galaxies with
central I band surface brightnesses 10\% of the night sky
brightness.

Our basic wavelength calibration procedure consists of fitting the centers of emission lines
produced by our Fe-cathode NeAr hollow cathode lamp with a low order polynomial, once for
each fiber, following the standard procedures in {\bf dofibers}. Typical
scatter about these fits is 0.06 {\rm \AA}, consistent with the error in
determining individual line centers. A single set of (typically) five
calibration exposures is taken in the late afternoon, with eight NeAr lamps
illuminating panels on the MMT's building shutters.

\begin{figure}
\epsscale{1}     
\plotone{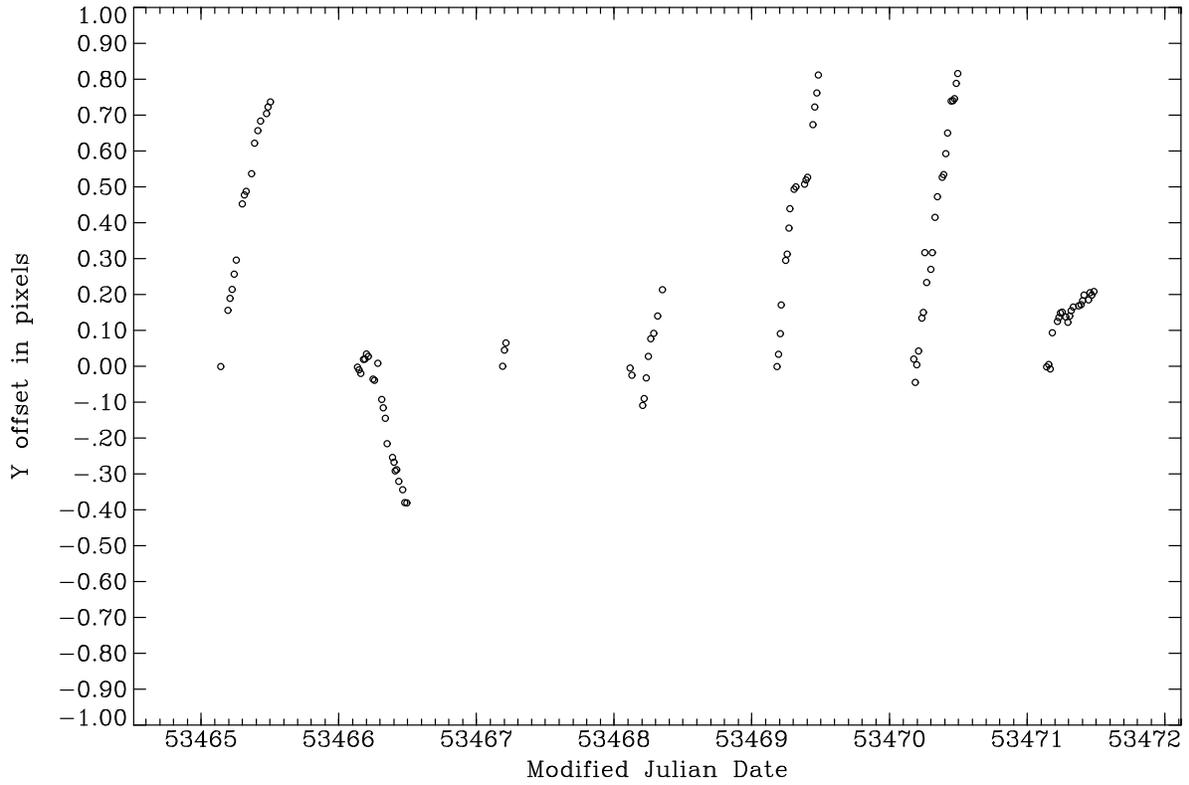}
\caption{The image motion (shift) in the dispersion direction for the week following 5 April 2005. The shifts are calculated by image correlation
and set to zero at the beginning of each night, as described in the text.\label{weekshift}}      
\end{figure}

The next step in the procedure corrects for thermally induced spectral shifts.
Purely linear image shifts can be removed easily. Figure \ref{weekshift}
shows the image motion in the dispersion direction for a typical week of
observing in April 2005. These shifts are obtained by cross correlating a
1000 {\rm \AA} wide image section centered near the $\lambda$8399{\rm \AA}
night sky line taken from the first on-sky CCD frame taken each night; thus
the drift shown in figure \ref{weekshift} is reset to zero each night. 
Typical nightly image motion is less than one pixel. The image motions in the direction
perpendicular to the dispersion show similar patterns and extents; the two
components of motion are not strongly correlated.
                                                                                           
\begin{figure}      
\epsscale{1}
\plotone{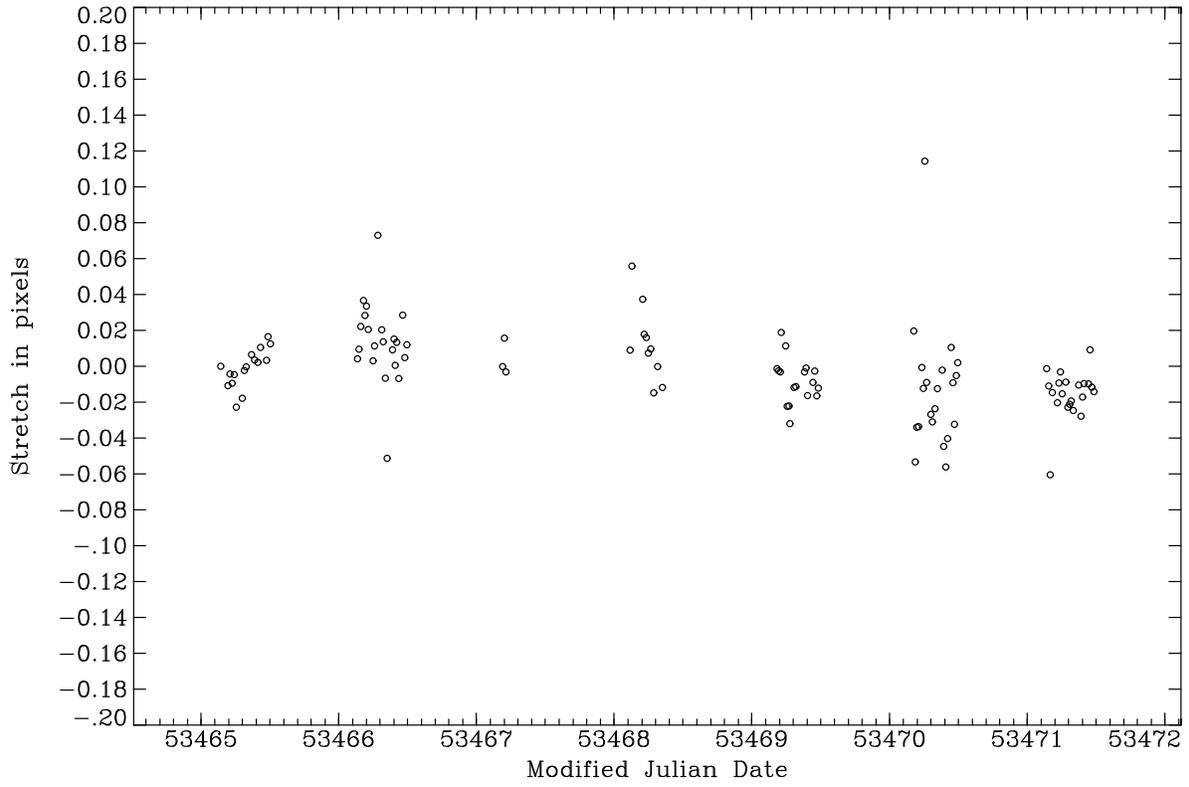}
\caption{The stretch distortion in the dispersion direction for the week following 5 April 2005. The stretch is calculated by the difference in the red and blue image correlation shifts, and is set to zero at the beginning of each night, as described in the text.\label{weekstretch}}
\end{figure}

Any non-linear shifts would appear to first order as stretches or rotations. Figure
\ref{weekstretch} shows the differences in the shift as measured by cross
correlation of image segments centered around $\lambda$8399 {\rm \AA} and
around $\lambda$5577 {\rm \AA} for the same week in April 2005 shown in Figure
\ref{weekshift}.  Notice that the y axis scale has been expanded by a factor of
five. The results are consistent with zero distortion within the measurement error of
0.02 pixels. A similar analysis shows that image rotations are also
consistent with zero.  A systematic change in the image scale of one part in 100,000
over twelve hours would result in a change in the measured distance of 0.02
pixels between $\lambda$8399 and $\lambda$5577. This is the level of the
scatter in figure \ref{weekstretch}, and represents a lower
limit on the image stability.

While the shifts measured by direct image correlation could put all on-sky
images onto the same system before the extraction of the 1-D spectra, this procedure
would not solve the problem of putting the on-sky images onto the same system
as the NeArFe calibration spectra. We choose to make the correction after the
(curved) 1-D spectra have been extracted, but before they are rebinned into
wavelength space. Compared with shifting before extracting this correction induces an
error which is proportional to the sine of the difference in the angle of
curvature of the spectra between the NeArFe spectrum and the on-sky spectrum,
at each fixed pixel position. For image shifts of $\sim$one pixel, even for
the most highly curved spectra, this error is negligible compared with the other
sources of error.

We use the standard {\bf apextract} techniques, determining the shift
perpendicular to the dispersion direction by image correlation with the flat
field image (which must be taken at nearly the same time as the NeArFe
calibrations). After extracting the 300 spectra from the image we simply
measure the centroid of the $\lambda$5577 night sky line, in pixel space, and
calculate (for each spectrum) the shift, in pixels between that position and
its expected position, as determined from the (300) NeAr solutions. We
assign the median shift to be the shift in the dispersion direction for the
entire image, and modify the wavelength polynomials accordingly. Finally, we
rebin the spectra into equal wavelength pixels.

\begin{figure}      
\epsscale{1}
\plotone{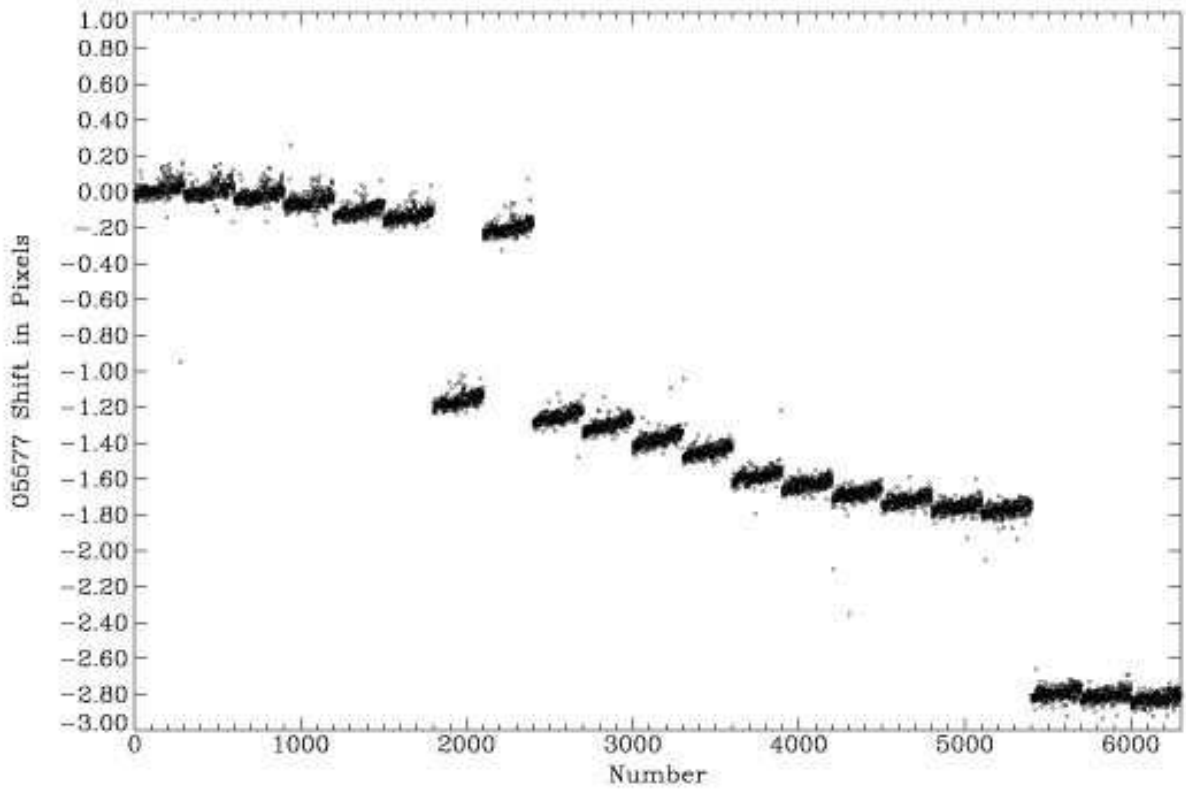}
\caption{The measured position of the O[I] 5577 night sky line for each spectrum measured on 10 April 2005, after extraction into 1-D pixel space, but before the shift correction has been applied.  The outliers in the plot are bad measurements due to cosmic ray hits near the sky line.  No image combination to reject cosmic rays was performed to maintain high time resolution.\label{5577}}
\end{figure}

Figure \ref{5577} shows the pixel shifts for each spectrum on the night of 10 April 2005, which is broadly
typical. Each point represents one spectrum, and each set of 300 points
represents one CCD frame. The spectra from each frame are plotted in order,
the time between frames is not shown, but can be seen in figure
\ref{weekshift} (the night in figure \ref{5577} is the second night from the right). Three things
are apparent in figure \ref{5577}.
(1) The shifts in the position of $\lambda$5577 are large when compared with other
effects. (2) The shifts change smoothly, but with occasional discontinuities
resulting from {\bf apextract's} choice of zeropoint. The discontinuities
demonstrate clearly that the shifts must be corrected before spectra may be
combined. (3) Finally, each set of 300 spectra, from a single frame, shows small,
systematic differences in the shift as a function of position in the
frame. The amplitude of this shift, from top to bottom, is similar to the
scatter in the measurement of a single line center. The origin of this shift
is unknown, but it is not due to systematic errors in the wavelength
polynomial. We currently do not correct for this small effect.

\begin{figure}      
\epsscale{1}
\plotone{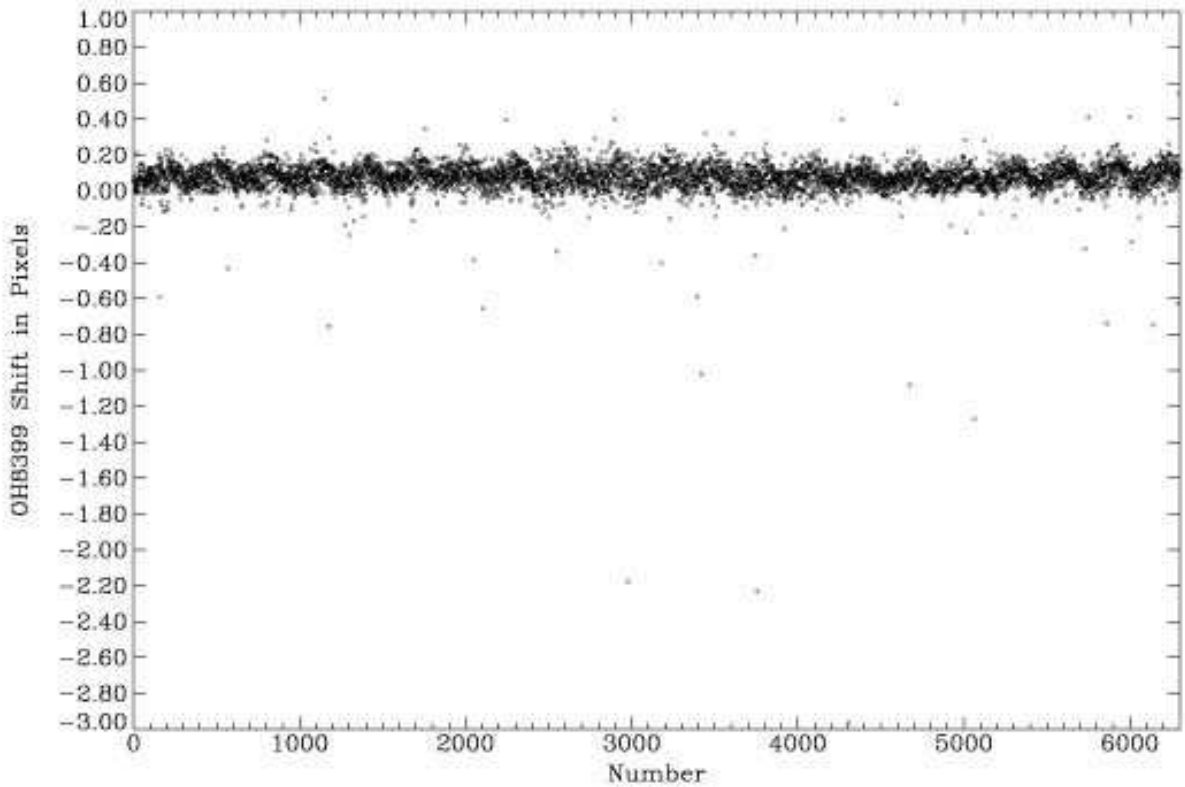}
\caption{The measured position of the OH 8399 night sky line for each spectrum measured on 10 April 2005, after all calibration steps have been taken.
The outliers in the plot are bad measurements due to cosmic ray hits near the sky line.  No image combination to reject cosmic rays was performed to maintain high time resolution.\label{8399}}
\end{figure}

Figure \ref{8399} shows the final measured positions for the
$\lambda$8399 {\rm \AA} night sky line after the spectra from figure \ref{5577}
have been wavelength calibrated and rebinned into 1-{\rm \AA}-wide equal
wavelength bins. The scatter about the expected value is 0.092 {\rm \AA}, which includes
the small systematic errors seen in Figure \ref{5577}.
The offset of 0.075 {\rm \AA} is approximately constant throughout the night, and may be
attributed to errors in the calibration polynomial as there is no apparent drift.

\subsection{\label{skysub} Sky Subtraction}

The removal of the night sky spectrum from observations of faint objects
made with fiber optic spectrographs has long been viewed as problematic
(e.g. \cite{1992MNRAS.257....1W}, \cite{1994PASP..106.1157L},
\cite{1998ASPC..152...50W}).  Recently methods have been developed which fit
local variations from the mean sky due to changes in fiber transmission, focus,
etc. by means of eigenvector techniques \cite{2000ApJ...533L.183K}, and
B-splines \cite{2001astro.ph..9164M}.
The stability of Hectospec's bench spectrograph allows us to reach
near Poisson limited sky subtraction without having to resort to the use of
eigenvectors or B-splines.

Typically 30 Hectospec fibers (10\% of the total) are allocated to blank sky
observations.  We reduce the data for each of the two CCD chips separately because 
combining sky spectra from the two
chips degrades performance.  The point spread functions of the two CCDs differ by
$\sim$0.2 pixels FWHM, apparently due to differences in the CCD charge diffusion.

We check for objects contaminating the sky spectra
in two ways.  (1) We divide each sky spectrum by the mean of all the sky
spectra.  When the RMS pixel value exceeds a threshold the sky spectrum is 
considered contaminated, and is rejected.
(2) We subtract the (revised) mean sky spectrum from each remaining sky
spectrum, then correlate the residuals against our absorption and emission
line galaxy templates; spectra which yield significant redshifts
(significance is determined by the \cite{1979AJ.....84.1511T} $r$ statistic)
are contaminated by galaxian or stellar light, and are rejected.  For a typical
sample, taken in March and April 2005, 181 of 972 sky spectra
($\sim$19\%) were found to be contaminated by this test.

We subtract the sky from each object spectrum by averaging the six closest (clean) sky spectra.
We rescale this average so that the flux in the OH 8399\AA\ line
matches the flux in the object spectrum, and then simply subtract the rescaled sky spectrum from the
object spectrum.
For galaxies with R$\sim$20 this procedure works very well.  We test
the limits of the technique by applying it to the measurement of
substantially fainter spectra.

\begin{figure}      
\plotone{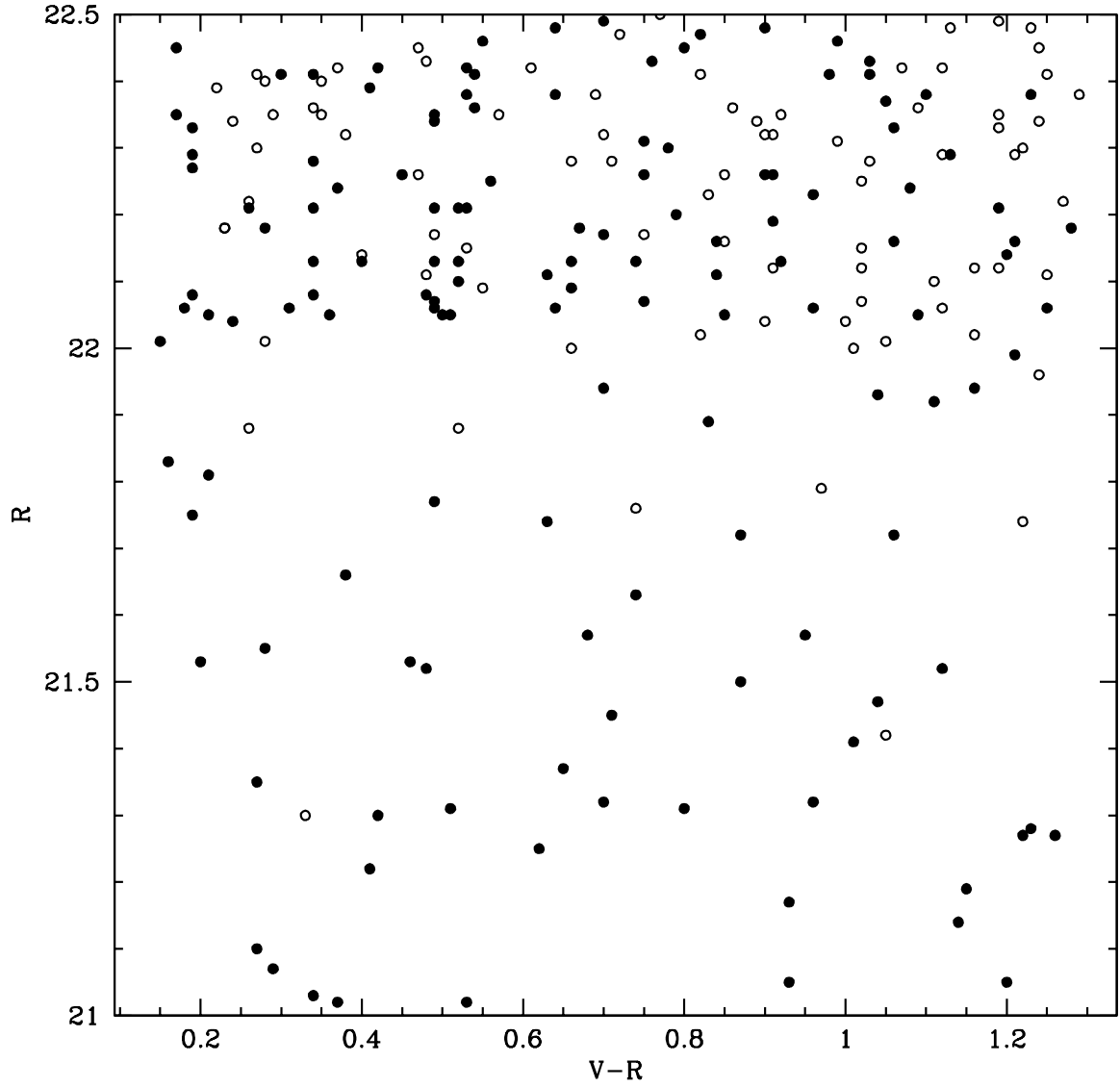}
\caption{\label{magcolor}
The distribution in $V-R$ color and isophotal R magnitude of the
objects used to test the sky subtraction techniques; filled circles 
represent objects with reliable redshifts after 280 minutes 
exposure.}
\end{figure}

We chose 222 objects from a deep galaxy survey \citep{dell05} that were evenly
distributed in $V-R$ color between $0.0 < V-R < 1.3$.   One third
were evenly divided in bright magnitude $21 < R < 22$ and two thirds were
evenly divided in faint magnitude $22 < R < 22.5$; figure \ref{magcolor}
shows the distribution of objects.  
These objects were observed in two different Hectospec configurations before
and after transit.  The first observation had 160
minutes exposure and the second 120 minutes exposure.  The two data sets were reduced
separately, then summed to provide the final spectra.  We obtained reliable
redshifts for 137 of these objects; the remainder had insufficient signal
to noise to obtain redshifts. The solid dots in figure \ref{magcolor} show 
the galaxies with redshifts; they tend to be brighter and bluer.
                                                                       
\begin{figure}      
\plotone{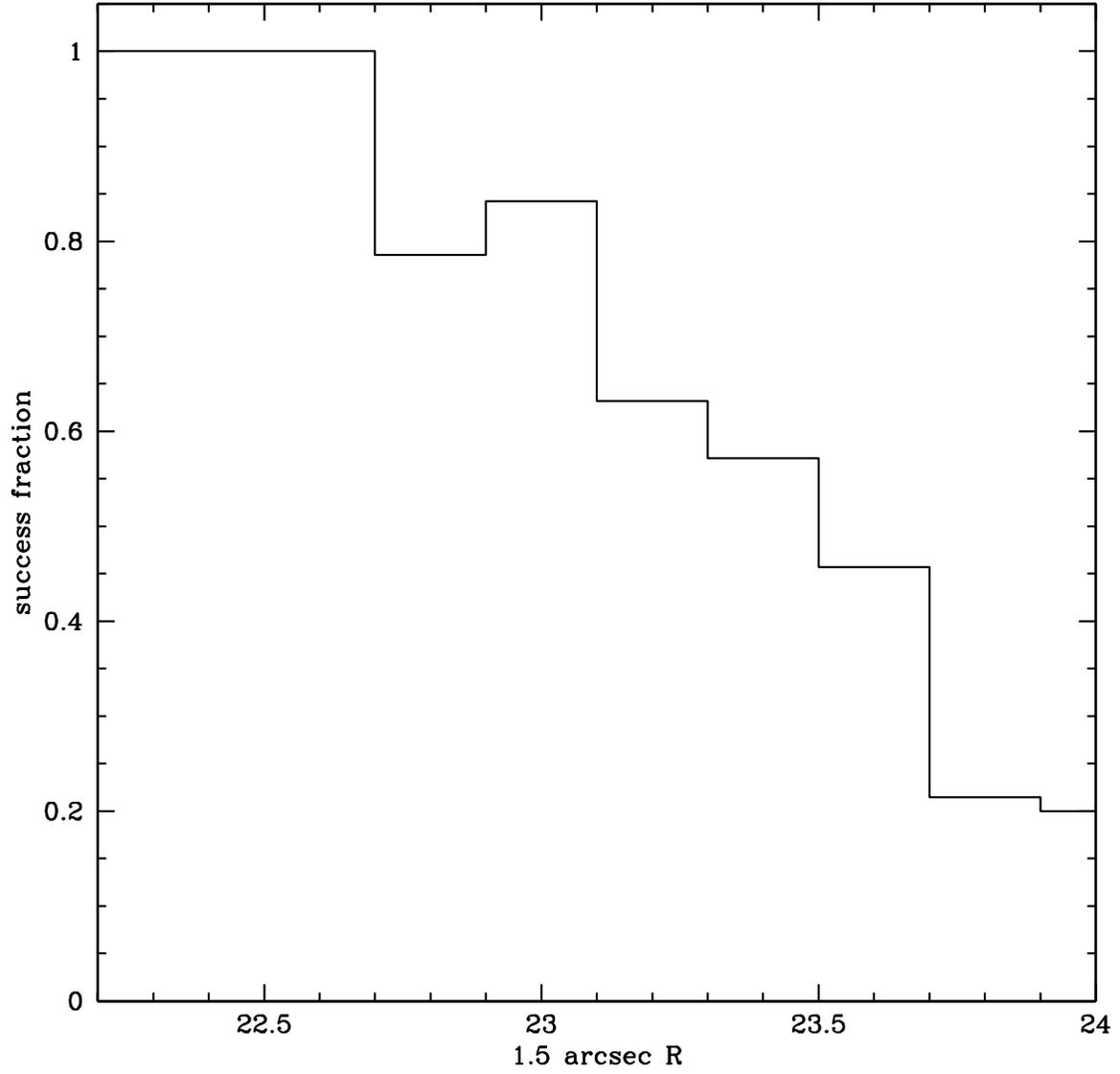}
\caption{\label{successrat}
The fraction of objects which have successful redshift measurements
as a function of the light down the fiber, a $1.5\arcsec$ aperture.
All of the lowest central surface brightness objects with successful
redshifts have strong emission lines.}
\end{figure}

Figure \ref{successrat} shows the success rate as a function of the
1.5${\arcsec}$ diameter aperture magnitude, which corresponds to the light
down the fiber.  The success rate falls rapidly for objects fainter then
R$_{1.5\arcsec}$=23.0; essentially all of the lower central surface
brightness objects with reliable redshifts are emission line objects.
                                                                      
\begin{figure}
\scalebox{.75}{\includegraphics[angle=-90]{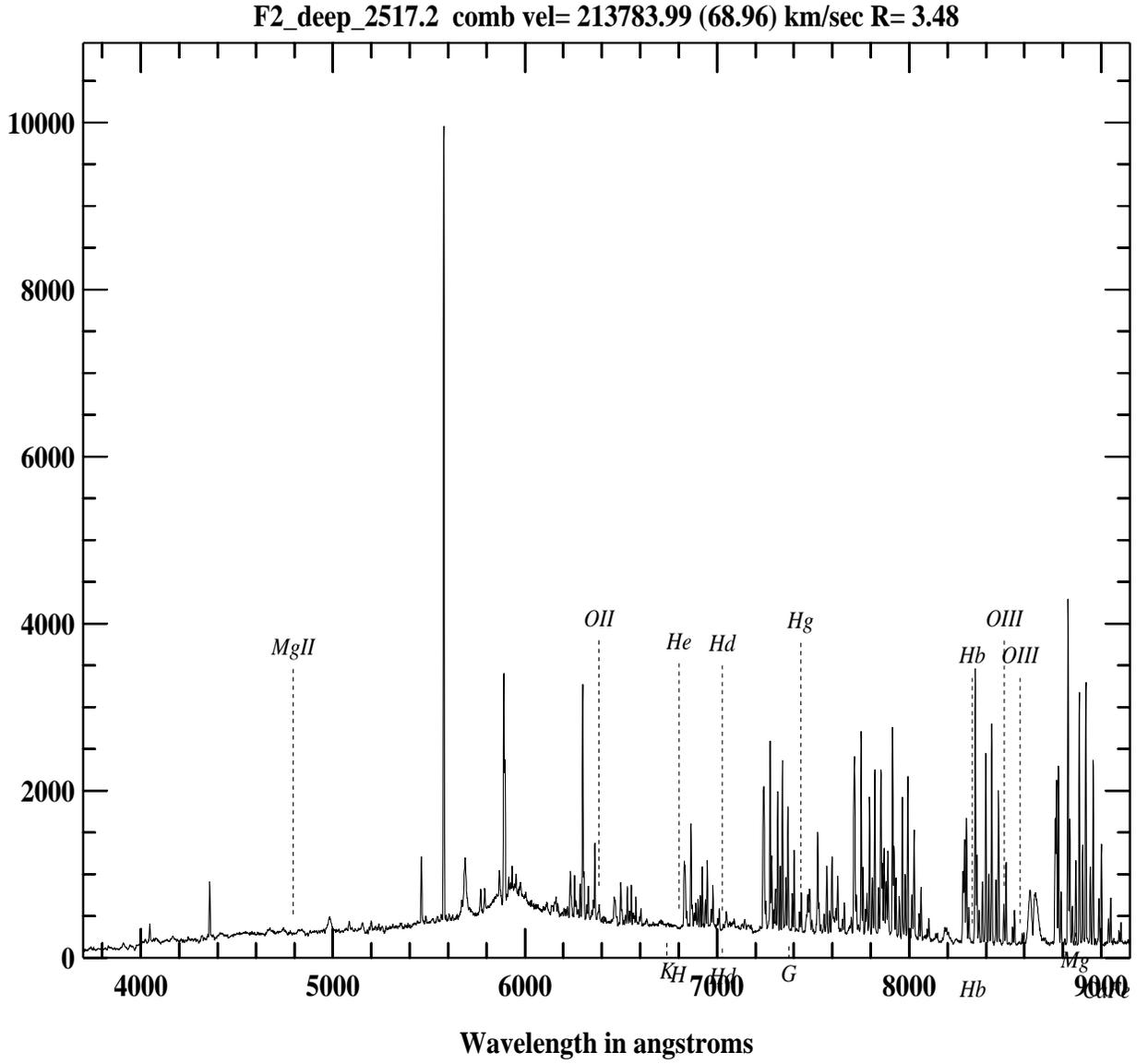}}      
\caption{\label{deepsky}
The spectrum of an R=22.06 galaxy with a reliable absorption line redshift
of 0.71, before sky subtraction.}
\end{figure}

\begin{figure}
 \scalebox{.75}{\includegraphics[angle=-90]{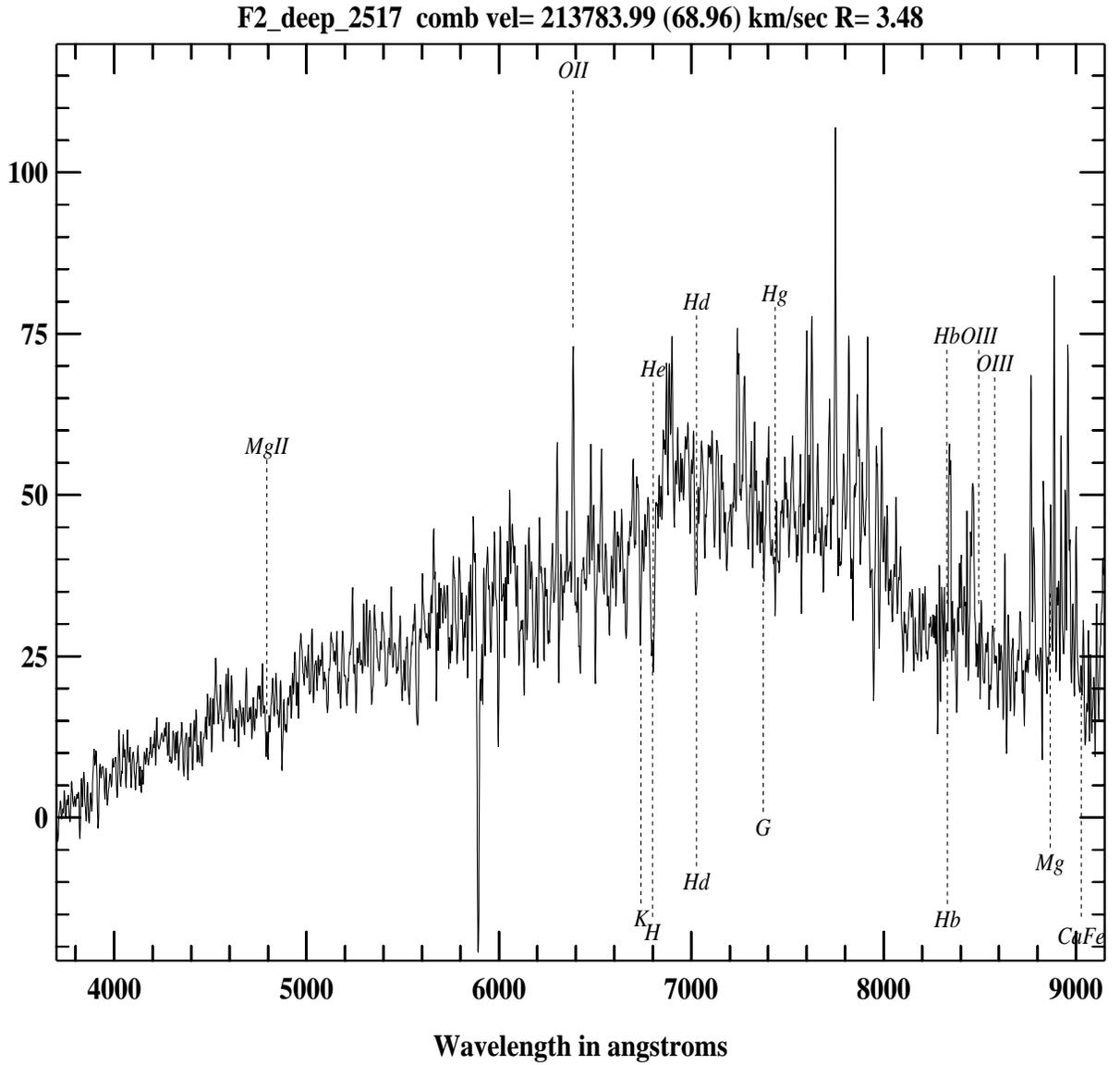}}    
\caption{\label{deepobj}
The spectrum of an R=22.06 galaxy with a reliable absorption line redshift
of 0.71, after sky subtraction.  The 1.5$^{\prime\prime}$ diameter aperture
magnitude is 22.97. The sky subtraction (see text) is effectively limited by
Poisson noise.}
\end{figure}

Figure \ref{deepsky} shows a 280 minute exposure spectrum of a
galaxy at z=0.71 having with R=22.0 and R$_{1.5\arcsec}$=22.97 prior to sky subtraction.
Figure \ref{deepobj} shows the spectrum after sky subtraction and after
application of an 8 \AA\ smoothing window. 
The CaII K line for this galaxy is detected in absorption at $\sim$1.5 $\sigma$ significance. 
There are about 450 independant 10 pixel wide regions in
this spectrum, we would expect about 30 of these to have positive
fluctuations greater than the size of the K line, and 30 to have negative
fluctuations.  This behavior is approximately what we see in this spectrum, taking
into account that the intensity of the sky radiation (and thus the size of
the statistical fluctuations) is in some places substantially higher than
at the location of the K line.

The coincidence of several features allows a reliable redshift to be
determined in this case, but the existence of many similar sized noise
spikes suggests that the spectrum of this R$\sim$22 galaxy is near the limit.  Were this object at z=1.0 the
light entering the fiber would be reduced by half and the spectral lines would
be redshifted into the OH night sky line forest.  With perfect sky
subtraction the K line would be about a 0.3 $\sigma$ effect; redshift
determinations are impossible at these levels.  While in theory
substantially longer exposures could be made to measure higher redshift
objects, in practice it makes more sense to use higher resolution to
isolate the night sky features \citep{2003SPIE.4834..161D}.  A 600 line
grating blazed in the red for the Hectospec will see first light during the
second half of 2005.

\section{\label{performance} Typical Hectospec Performance in Survey Mode}

\begin{figure}
\scalebox{.75}{\includegraphics[angle=-90]{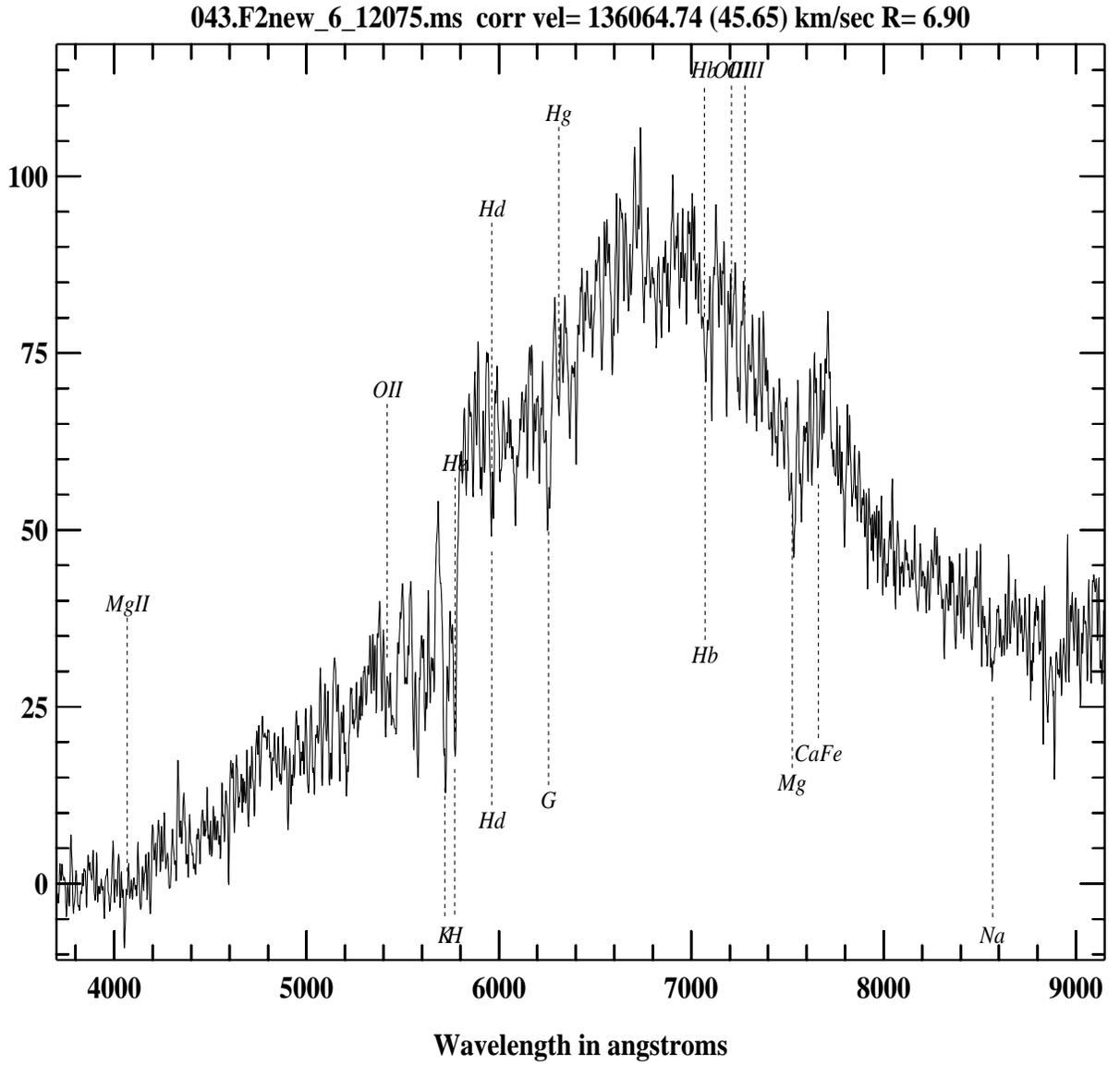}}      
\caption{\label{typabs}
The spectrum of a typical absorption line galaxy, from a one hour 
exposure with the moon up.  The object has an isophotal R 
magnitude of 19.00,
a 1.5$^{\arcsec}$ magnitude of 21.23, and a redshift of 0.45.}
\end{figure}

\begin{figure}      
\scalebox{.75}{\includegraphics[angle=-90]{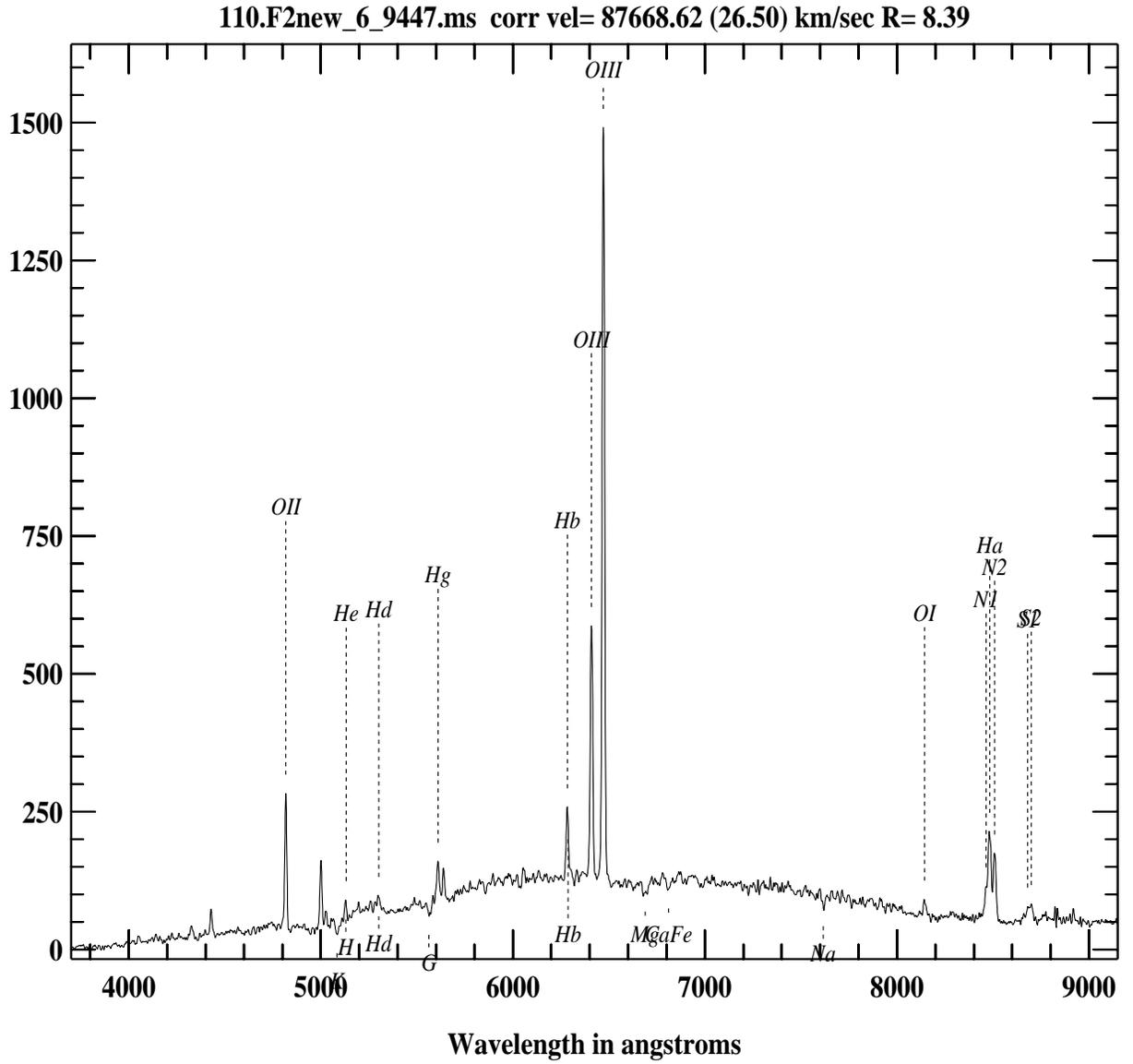}}
\caption{\label{typem}
The spectrum of a typical emission line galaxy, from a one hour 
exposure with the moon up.  The object has an isophotal R 
magnitude of 19.12,
a 1.5$^{\arcsec}$ magnitude of 20.31, and a redshift of 0.29.}
\end{figure}

Hectospec is frequently used to observe objects with $R<21.0$ 
with exposures of 30 to 120 minutes. Often observations are made
with partial moon.  Figures \ref{typabs} and \ref{typem} show
representative 60 minute exposures, taken with the moon up, six
days past the new moon.
Redshifts are determined using the methods described by
\citet{kurtz98}, hereafter KM98.  New templates, using
objects with z$\sim$0.3 were created for the correlation
analysis, as described in KM98.

\begin{figure}      
\plotone{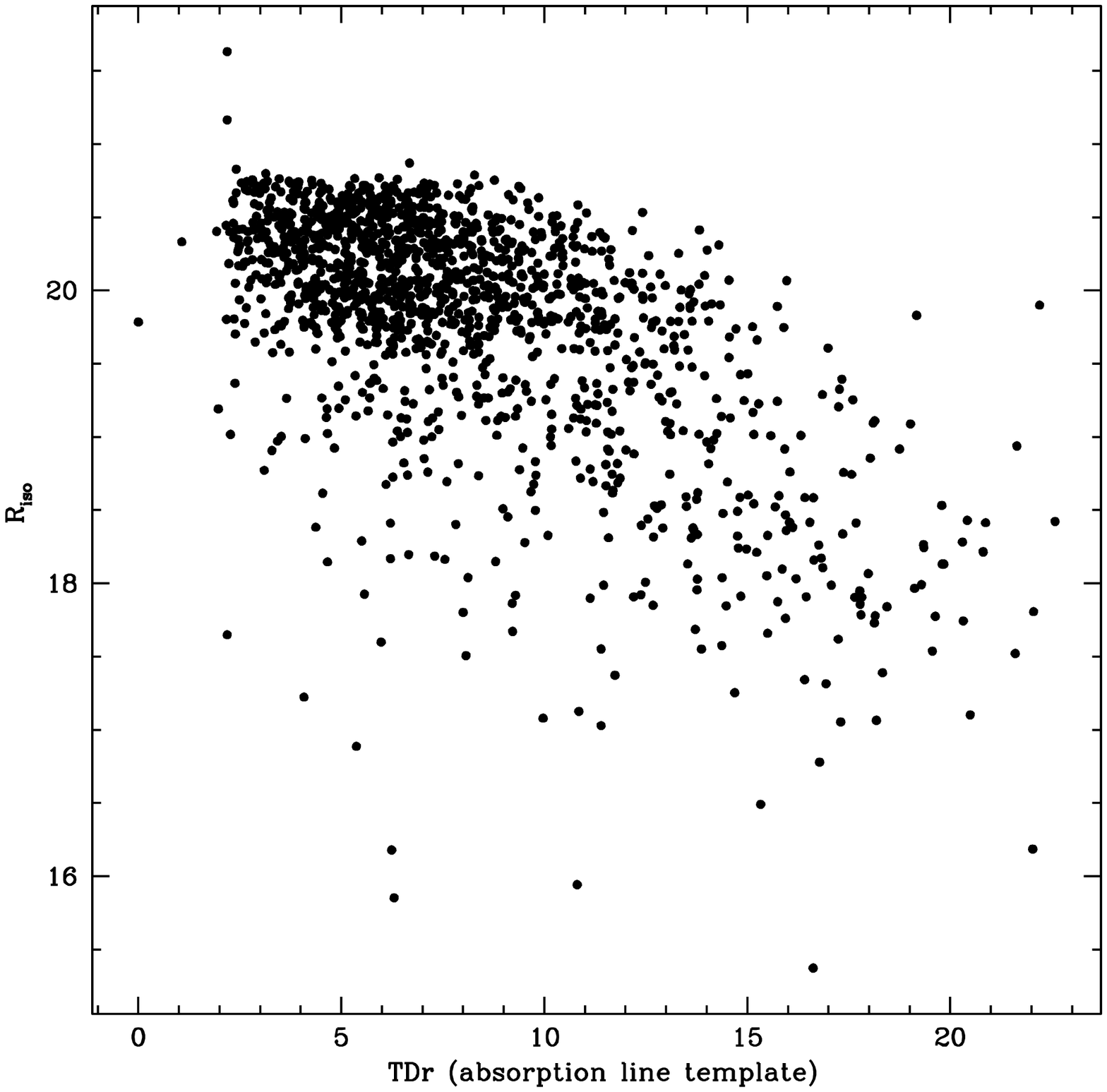}
\caption{\label{isoVSr}
The isophotal R magnitude vs. the quality of the redshift
as measured by the \citet{1979AJ.....84.1511T} $r$ statistic
for 1455 one hour exposures of absorption line galaxies.}
\end{figure}

Figure \ref{isoVSr} shows the relation between isophotal R magnitude
and the \citet{1979AJ.....84.1511T} $r$ statistic, a measure of
redshift quality, for 1455 absorption line galaxies observed with one
hour exposures (some with moon) in March and April 2005 as part of the
gravitational lensing survey \citep{gel05}.  All these spectra yielded reliable redshifts.  
There is little correlation between the magnitude and the quality of the
redshift.

\begin{figure}      
\plotone{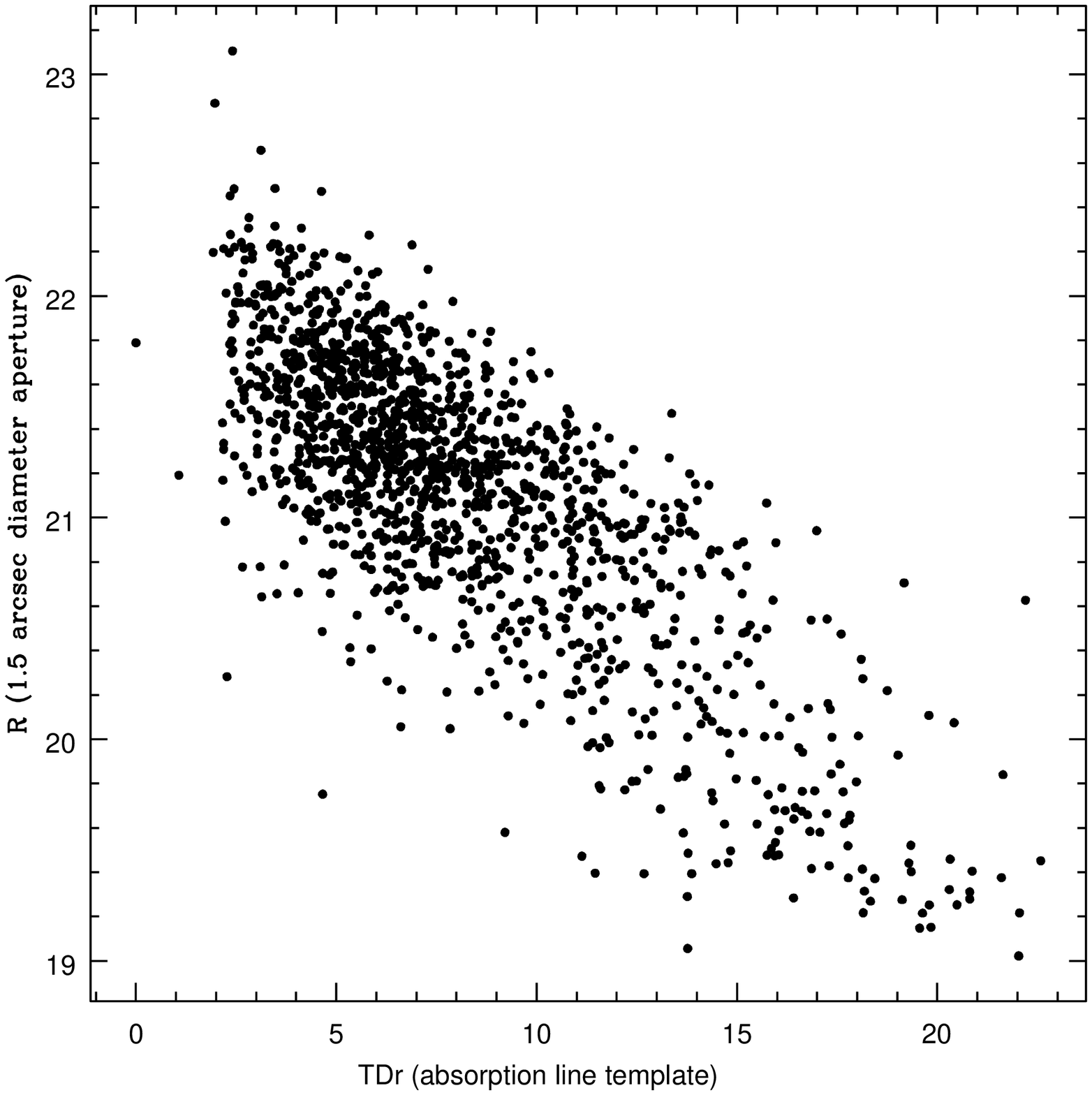}
\caption{\label{aperVSr}
The 1.5$^{\arcsec}$ aperture R magnitude vs. the quality of the redshift
as measured by the \citet{1979AJ.....84.1511T} $r$ statistic
for 1455 one hour exposures of absorption line galaxies.}
\end{figure}

Figure \ref{aperVSr} shows the same 1455 absorption line galaxies, but
with the 1.5$^{\arcsec}$ aperture magnitude, which represents the light
down the fiber, vs. the $r$ statistic.  As expected there is now a strong correlation.
The Mt. Hopkins R band dark sky brightness is
typically 20.0 in a 1.5$^{\arcsec}$ aperture
\citep{2000PASP..112..566M,2003A&A...400.1183P}. The Hectospec
routinely obtains absorption line redshifts for objects at 20\%\ of the sky in one
hour exposures.

Following KM98 we use 332 objects with duplicate spectra to estimate
the error in a redshift measurement as a function of the $r$
statistic.  We find that the median velocity error for absorption line
redshifts can described by $\Delta v_{abs} = 130/(1+r) + 5$ km s$^{-1}$
while the emission line redshift errors can be described by $\Delta
v_{em} = 100/(1+r) + 5$ km s$^{-1}$.  The mean errors are about twice
these, and are well represented by the normal {\bf xcsao} error
estimator (KM98).

\section{Discussion and Conclusions}

Hectospec is a powerful, wide-field spectrograph that makes
excellent use of the converted MMT's 1$^{\circ}$ diameter and 6.5 meter
aperture. It reaches to R=21.5 with ease, and
the sky subtraction with standard data reduction allows quality spectroscopy
to R=22 with sufficient exposure. Fiber spectrographs have an unjustified
reputation in some quarters for compromised throughput that is probably due to
design shortcomings in the fiber run between the focal plane and the
spectrograph and in the design of bench spectrographs. Hectospec's peak system
throughput of 10\% including aperture losses, and 17\% correcting for aperture
losses, is quite competitive with slit spectrographs when its huge multiplex
advantage is taken into account. A high-throughput long-slit spectrograph like
the FAST spectrograph on the Whipple Observatory's 1.5m Tillinghast Reflector
\citep{fast} has a peak system throughput (correcting for aperture losses by
using a wide slit) that reaches 30\% with fresh mirror and corrector coatings,
but is more typically $\sim$25\% in average conditions.

Hectospec is a popular instrument at the MMT.  Users
respond favorably to the flexibility and efficiency of the queue observing mode.
We designed Hectospec to rapidly survey normal intermediate redshift galaxies 
and high redshift active galaxies identified through by space X-ray and infrared
observatories, and these are active areas of research with Hectospec.  A wide 
variety of other programs are underway, including several stellar programs.
Because Hectospec can offer effective sky subtraction to R$\sim$22, essentially
limited by Poisson noise, it is a surprisingly versatile general-purpose spectrograph.

We are in the process of building a complementary optical imaging
spectrograph for the MMT (Binospec, \citet{fab03}).  A wide-field imaging spectrograph
will allow us to efficiently survey galaxy dynamics at intermediate redshifts and will
offer very high throughput (potentially twice Hectospec's) and 
even more precise sky subtraction for very faint objects.  However,
Binospec's advantages come at a price: its field of view is only 8\% of Hectospec's, and
with likely slitlet lengths of 10$^{\prime\prime}$ to 12$^{\prime\prime}$, 
Binospec will offer $\sim$50\% of Hectospec's multiplex
advantage.  The Hectospec will remain the instrument of choice for
large surveys to depths of R$\sim$22.

\acknowledgments

We are grateful for the contributions of the entire MMT instrument team,
students, and of the MMT and Whipple Observatory staff. We thank Steve Amato,
David Becker, Kevin Bennett, David Bosworth, David Boyd, Daniel Blanco, William
Brymer, David Caldwell, Richard Cho, Shawn Callahan, Florine Collette, Dylan Curley, Emilio
Falco, Craig Foltz, Art Gentile, Everett Johnston, Sheila Kannappan, Frank
Licata, Steve Nichols, Dale Noll, Ricardo Ortiz, Tim Pickering, Frank Rivera,
Phil Ritz, Rachel Roll, Cory Sassaman, Gary Schmidt, Dennis Smith, Ken Van Horn, David
Weaver, Grant Williams and J.T. Williams. We thank the Hectospec robot
operators: Perry Berlind and Mike Calkins, and the MMT operators: Mike
Alegria, Alejandra Milone, and John McAfee. We are grateful to the efforts of
the Harvard College Observatory Model Shop led by Larry Knowles. We thank
Robert Dew and Diane Nutter of Cleveland Crystals and Terry Facey, formerly of
Goodrich Optical Systems.

Facilities: \facility{MMT}.

\clearpage

\clearpage

\clearpage

\clearpage

\clearpage

\clearpage

\clearpage

\clearpage

\clearpage

\clearpage

\clearpage

\clearpage

\begin{center}
\begin{deluxetable}{lllll}
\tabletypesize{\scriptsize}
\tablecaption{Fiber Positioner Move Times\label{tblmovet}}
\tablewidth{0pt}
\tablehead{
\colhead{Move Parameter} & \colhead{X \& Y Axes} & \colhead{Z Axis} & 
\colhead{$\Theta$ \& $\Phi$ Axes} & \colhead{Gripper} 
}
\startdata
S-Curve Time          &  0.030 s           & 0.015 s         & 0.020 s       & 0.003 s           \\
Acceleration Time     &  0.100 s           & 0.030 s         & 0.070 s       & 0.050 s           \\
Velocity              &  700 mm s$^{-1}$   & 120 mm s$^{-1}$ & 7 mm s$^{-1}$ & 938 steps s$^{-1}$\\
Settling Time\tablenotemark{a}     & 0.100 s & 0.050 s       & 0.100 s       &  0 s              \\   
Typical Move          &  150 mm            &  10 mm          & 2 mm          &  38 steps         \\
Typical Move Time\tablenotemark{b} &  0.467 s\tablenotemark{c}& 0.193 s & 0.426 s\tablenotemark{c}& 0.040 s \\
\enddata

\tablenotetext{a}{Time for servo system to reach position error less than 0.005 mm.} 
\tablenotetext{b}{We ignore the small motion during the S-curve acceleration, so these times are a slight overestimate.}
\tablenotetext{c}{The XY and $\Theta\Phi$ moves are executed simultameously, so the longer time will dominate.}
\end{deluxetable}

\end{center}

\begin{center}
\begin{deluxetable}{ll}
\tabletypesize{\scriptsize}
\tablecaption{Fiber Positioning Error Sources\label{poserr}}
\tablewidth{0pt}
\tablehead{
\colhead{Error Source} & \colhead{Error Magnitude (mm)} 
}
\startdata
Servo Loop Closure                                     &  0.005\\
Axis Calibration onto Orthogonal Cartesian Coordinates &  0.008\\
Button Release Error                                   &  0.008\\
Calibration of Fiber Within Button                     &  0.005\\
Focal Surface Flexure Correction                       &  0.005\tablenotemark{a}\\
Flexure of Guide Probes During Observation             &  0.005\tablenotemark{a}\\
Total Positioning Error Added in Quadrature            &  0.015\\
Total Positioning Error Summed                         &  0.036\\

\enddata

\tablenotetext{a}{Estimated}

\end{deluxetable}
\end{center}

\begin{center}
\begin{deluxetable}{ll}
\tabletypesize{\scriptsize}
\tablecaption{Fiber Positioning Reconfiguration Overhead\label{overhead}}
\tablewidth{0pt}
\tablehead{
\colhead{Activity} & \colhead{Elapsed Time (s)} 
}
\startdata
Readout CCD and Request Slew &  60 s\\
Slew Telescope to Zenith     &  60 s\\
Reconfigure Fibers           & 320 s\\
Move Guider Probes           &  30 s\\
Slew to WFS Star             &  60 s\\
Acquire WFS and Correct      & 300 s\\
Slew to Field and Track      &  30 s\\
Setup on Guide Stars         & 180 s\\
Total                        &1040 s\\
\enddata

\tablenotetext{a}{Estimated}

\end{deluxetable}

\end{center}


\begin{thebibliography}{}

\bibitem[Barden et al.(1993)]{Ba93} Barden et al. 1993, ASP 
Conf.~Ser.~37: Fiber Optics in Astronomy II, 152, 185 
\bibitem [Blanco et al.(2004) ]{blanco04} Blanco et al. 2004, \procspie, 5489, 300
\bibitem [Callahan et al.(2004) ]{cal04} Callahan, S., Cuerdan, B., Fabricant, D., \& Martin, H. 2004, \procspie, 5495, 228
\bibitem[Davis et al.(2003)]{2003SPIE.4834..161D} Davis, M., et al.\ 2003, 
\procspie, 4834, 161
\bibitem [Dell'Antonio et al.(2005) ]{dell05} Dell'Antonio, I. 2005, private communication 
\bibitem [Epps \& Vogt(1993) ]{hires93} Epps, H. \& Vogt, S., \ao, 32, 6272
\bibitem [Fabricant et al.(1998a) ]{fast} Fabricant, D. et al. 1998, \pasp, 110, 79
\bibitem [Fabricant et al.(1998b) ]{fab98} Fabricant, D. et al. 1998, \procspie, 3355, 285
\bibitem [Fabricant et al.(2003) ]{fab03} Fabricant, D. et al. 2003, \procspie, 4841, 134
\bibitem [Fabricant et al.(2004) ]{fab04} Fabricant, D. et al. 2004, \procspie, 5492, 767
\bibitem [Fabricant, Hertz \& Szentgyorgyi(1994) ]{fab94}Fabricant, D., Hertz, E., \& Szentgyorgyi, A. 1994, \procspie, 2198, 251
\bibitem [Fata, Kradinov \& Fabricant(2004) ]{fata04} Fata, R., Kradinov, V., \& Fabricant, D. et al. 2004, \procspie, 5492, 553
\bibitem [Fata \& Fabricant(1998) ]{fata98} Fata, R \& Fabricant, D. 1998, \procspie, 3355, 275
\bibitem [Fata \& Fabricant(1994) ]{fata94} Fata, R \& Fabricant, D. 1994, \procspie, 2199, 580
\bibitem [Fata \& Fabricant(1993) ]{fata93} Fata, R \& Fabricant, D. 1993, \procspie, 1998, 32
\bibitem [Geary (2000) ]{geary00}Geary, J. 2000, {\it in Further Developments in Scientific Optical Imaging}, ed. Denton, M. B.,
(Cambridge, U.K.: Roy. Soc. Chem.)
\bibitem[Geller et al.(2005) ]{gel05} Geller, M.~J. 2005, to be submitted to \apjl
\bibitem[Kurtz \& Mink(1998) ]{kurtz98} Kurtz, M.~J., \& Mink, D.~J.\ 1998, \pasp, 110, 934 
\bibitem[Kurtz \& Mink(2000)]{2000ApJ...533L.183K} Kurtz, M.~J., \& Mink, 
D.~J.\ 2000, \apjl, 533, L183 
\bibitem[Lissandrini et al.(1994)]{1994PASP..106.1157L} Lissandrini, C., 
Cristiani, S., \& La Franca, F.\ 1994, \pasp, 106, 1157 
\bibitem[Marinoni et al.(2001)]{2001astro.ph..9164M} Marinoni, C., Davis, 
M., Coil, A.~L., \& Finkbeiner, D.\ 2001, in Where's the Matter?,
Proceeding of the 3rd Marseille Cosmology Conference, eds. L. Tresse and
M. Treyer, p. 118, also astro-ph/0109164
\bibitem[Massey \& Foltz(2000)]{2000PASP..112..566M} Massey, P., \& Foltz, 
C.~B.\ 2000, \pasp, 112, 566 
\bibitem[Mink et al.(2005) ]{mink} Mink, D.~M., Wyatt, W.~F., Roll, J.~B., Tokarz, S.~B., Conroy, M.~A., Caldwell, N., Kurtz, M.~J., and
Geller, M.~J., Astronomical Data Analysis Software and Systems XIV, ASP Conference Series, in press.
\bibitem [Parry \& Gray(1986) ]{parry86} Parry, I \& Gray, P. 1986, \procspie, 627, 118
\bibitem[Patat(2003)]{2003A&A...400.1183P} Patat, F.\ 2003, \aap, 400, 1183 
\bibitem [Pickering, West \& Fabricant(2004) ]{pick04} Pickering, T., West, S., \& Fabricant, D. 2004, \procspie, 5489, 1041
\bibitem [Roll, Fabricant \& McLeod(1998) ]{roll98} Roll, J.B. Jr., Fabricant, D., \& McLeod, B. 1998, \procspie, 3355, 324
\bibitem[Roll(1996) ]{roll96} Roll, J.\ 1996, ASP Conf.~Ser.~101: Astronomical Data Analysis Software and Systems V, 101, 536 
\bibitem [Szentgyorgyi et al.(1998) ]{saint98} Szentgyorgyi, A. et al. 1998, \procspie, 3355, 242
\bibitem[Tokarz \& Roll(1997) ]{tokarz97} Tokarz, S.~P., \& Roll, J.\ 1997, ASP Conf.~Ser.~125: Astronomical Data Analysis Software and 
Systems VI, 125, 140 
\bibitem[Tonry \& Davis(1979)]{1979AJ.....84.1511T} Tonry, J., \& Davis, 
M.\ 1979, \aj, 84, 1511 
\bibitem[Valdes(1995) ]{valdes95} Valdes, F.\ 1995, ``Guide to the Multifiber Reduction Task DOFIBERS,'' IRAF On-line document;
http://iraf.noao.edu.
\bibitem[Valdes(1992) ]{valdes92} Valdes, F.\ 1992, ASP Conf.~Ser.~ 25: Astronomical Data Analysis Software and Systems I, 25, 417 
\bibitem[Valdes(1995)]{1995IRAF..........V} Valdes, F.\ 1995, ``Guide to
the Multifiber Reduction Task DOFIBERS,'' IRAF On-line document;
http://iraf.noao.edu.
\bibitem [Vogt et al.(1994) ]{vogt94} Vogt, S. et al. 1994, \procspie, 2198, 362
\bibitem[Watson et al.(1998)]{1998ASPC..152...50W} Watson, F., Offer, 
A.~R., Lewis, I.~J., Bailey, J.~A., \& Glazebrook, K.\ 1998, ASP 
Conf.~Ser.~152: Fiber Optics in Astronomy III, 152, 50 
\bibitem[Wyse \& Gilmore(1992)] {1992MNRAS.257....1W} Wyse, R.~F.~G., \& 
Gilmore, G.\ 1992, \mnras, 257, 1 
 
 
 



\end{thebibliography}
\end{document}